\documentstyle[twocolumn,aps,prd,floats,epsfig]{revtex}

\begin{document}

\draft

\wideabs{

\title{Search for the lepton-family-number nonconserving decay $\mu^+ \to e^+ \gamma$}

\author{
M.~Ahmed$^{c,}$\cite{byline1},
J.F.~Amann$^e$,
D.~Barlow$^{i,}$\cite{byline2},
K.~Black$^{e,}$\cite{byline3},
R.D.~Bolton$^e$,
M.L.~Brooks$^e$,
S.~Carius$^{e,}$\cite{byline4},
Y.K.~Chen$^{c,}$\cite{byline5},
A.~Chernyshev$^e$,
H.M.~Concannon$^j$,
M.D.~Cooper$^e$,
P.S.~Cooper$^b$,
J.~Crocker$^{a,}$\cite{byline6},
J.R.~Dittmann$^{j,}$\cite{byline7},
M.~Dzemidzic$^{c,}$\cite{byline8},
A.~Empl$^c$,
R.J.~Fisk$^j$,
E.~Fleet$^{l,}$\cite{byline9},
W.~Foreman$^e$,
C.A.~Gagliardi$^h$,
D.~Haim$^{l,}$\cite{byline10},
A.~Hallin$^{e,}$\cite{byline27},
C.M.~Hoffman$^e$,
G.E.~Hogan$^e$,
E.B.~Hughes$^{g,}$\cite{byline11},
E.V.~Hungerford~III$^c$,
C.C.H.~Jui$^{g,}$\cite{byline12},
G.J.~Kim$^h$,
J.E.~Knott$^d$,
D.D.~Koetke$^j$,
T.~Kozlowski$^e$,
M.A.~Kroupa$^{e,}$\cite{byline13},
A.R.~Kunselman$^m$,
K.A.~Lan$^c$,
V.~Laptev$^e$,
D.~Lee$^e$,
F.~Liu$^{h,}$\cite{byline14},
R.W.~Manweiler$^j$,
R.~Marshall$^k$,
B.W.~Mayes~II$^c$,
R.E.~Mischke$^e$,
B.M.K.~Nefkens$^i$,
L.M.~Nickerson$^{j,}$\cite{byline15},
P.M.~Nord$^j$,
M.A.~Oothoudt$^e$,
J.N.~Otis$^{g,}$\cite{byline16},
R.~Phelps$^{c,}$\cite{byline17},
L.E.~Piilonen$^l$,
C.~Pillai$^{e,i}$,
L.~Pinsky$^c$,
M.W.~Ritter$^{g,}$\cite{byline18},
C.~Smith$^{l,}$\cite{byline19},
T.D.S.~Stanislaus$^{e,j}$,
K.M.~Stantz$^{d,}$\cite{byline20},
J.J.~Szymanski$^e$,
L.~Tang$^{c,}$\cite{byline21},
W.B.~Tippens$^{k,i,}$\cite{byline22},
R.E.~Tribble$^h$,
X.L.~Tu$^{h,}$\cite{byline23},
L.A.~Van~Ausdeln$^{h,}$\cite{byline24},
W.H.~von~Witch$^{c,}$\cite{byline25},
D.~Whitehouse$^{e,}$\cite{byline26},
C.~Wilkinson$^e$,
B.~Wright$^k$,
S.C.~Wright$^a$,
Y.~Zhang$^{l,}$\cite{byline27},
and K.O.H.~Ziock$^{k,}$\cite{byline28}\\
(MEGA Collaboration)
}

\address{
$^a$University of Chicago, Chicago, IL 60637\\
$^b$Fermi National Accelerator Laboratory, Batavia, IL 60510\\
$^c$University of Houston, Houston, TX 77204\\
$^d$Indiana University, Bloomington, IN 46383\\
$^e$Los Alamos National Laboratory, Los Alamos, NM 87545\\
$^g$Stanford University, Stanford, CA 94305\\
$^h$Texas A \& M University, College Station, TX 77843\\
$^i$University of California Los Angeles, Los Angeles, CA 90095\\
$^j$Valparaiso University, Valparaiso, IN 46383\\
$^k$University of Virginia, Charlottesville, VA 22901\\
$^l$Virginia Polytechnic Institute and State University, Blacksburg, VA
24061\\
$^m$University of Wyoming, Laramie, WY 82071\\
}

\date{\today}

\maketitle

\begin{abstract}
The MEGA experiment, which searched for the muon- and electron-number
violating decay $\mu^+ \to e^+ \gamma$, is described.
The spectrometer system, the 
calibrations, the data taking procedures, the data analysis, and the sensitivity
of the experiment are discussed.  The most stringent upper limit on the 
branching ratio, ${\cal B}(\mu^+ \to e^+ \gamma) < 1.2\,\times\,10^{-11}$
with 90\% confidence, is derived from a likelihood analysis.
\end{abstract}


} % end wideabs here

\section{Introduction}

The standard $SU(3) \times SU(2) \times U(1)$ model of the strong and electroweak interactions
has proven to be remarkably successful in describing current experimental results, with only the
evidence for neutrino oscillations \cite{SNO} and the recent measurement of the muon
anomalous
magnetic
moment \cite{gminus2} falling outside its expectations.  The standard model is
nonetheless believed to be an effective
low-energy approximation of a more fundamental theory as it contains many free
parameters and unexplained symmetries.  Many
extensions to the standard model have been proposed.  Often these are motivated by attempts to
justify features, like parity violation or
lepton-family-number conservation, that are put in ``by hand" to explain experimental
data.

Essentially every extension of the standard model includes new heavy particles that mediate rare
decays that are otherwise forbidden.  The most sensitive decay mode varies by
model, so it is important to study a range of rare decays in the search for new physics.  The rare
decay $\mu \to e \gamma$ is the classic example of a reaction that would be allowed
except
for the separate conservation of muon and electron numbers within the standard model; in
fact, $\mu \to e \gamma$ is predicted to occur in most proposed extensions.  For example, it
has been shown that a broad range of grand unified supersymmetric theories predict that $\mu \to
e\gamma$ should occur with a branching ratio in the range 10$^{-10}$ to 10$^{-14}$
\cite{barbi95}.

The MEGA Collaboration was formed to search for the decay
$\mu \to e\gamma$ at the Los Alamos Meson Physics Facility (LAMPF).  The most sensitive
previous limit, ${\cal B}(\mu^+ \to e^+ \gamma) < 4.9\,\times\,10^{-11}$ (90\% C.L.) was
obtained with the Crystal Box detector \cite{bolton88}.  The MEGA experiment produced 
a new upper limit on the branching ratio, ${\cal B}(\mu^+ \to e^+ \gamma) <
1.2\,\times\,10^{-11}$ (90\% C.L.), as discussed briefly in Ref.\@ \cite{coop99}.  This
paper presents a more complete description of the experiment and its results.

\section{Experimental Design Philosophy}
\label{sect:design-philosophy}

The MEGA experiment was designed to find the decay $\mu^+ \to e^+\gamma$ in a
background-free environment if its branching ratio was greater than 10$^{-13}$.  The decay was
to be isolated from all backgrounds by its unique kinematic signature:  a muon at rest 
decaying into a time and spatially coincident photon and 
positron with equal and opposite momenta of 52.8 MeV/c.  The identification of the $\mu \to
e\gamma$ signal
relied on precise and accurate measurements of the vector momenta of the photon and the
positron at the muon
decay point and their relative times.

Accordingly, a high precision magnetic spectrometer system was constructed with two
distinct and separated parts:  (a) a
low-mass system of multiple-wire proportional chambers (MWPCs) to
track the positron orbits plus a series of plastic scintillators to determine the
end-time of the positron orbit, and
(b) pair spectrometers to detect the photons and determine their energy, propagation
direction, conversion time and location.  To extract the vector momentum of the photon, the
intersection
of the positron with the target was assumed to be the origin of the photon.  The replacement 
of the total absorption
calorimeter used in previous experiments by pair spectrometers was a trade-off of detection
efficiency for resolution, solid angle, and temporal stability.  To maximize the solid angle
acceptance, the photon system nearly surrounded the positron detection system.  The
magnetic field was produced by a superconducting solenoid, and the spectrometer system was
cylindrical in shape, with the cylinder axis parallel to the solenoid field.

The positron MWPCs were constructed with as little mass as possible to minimize
$dE/dx$ energy loss, annihilation in flight, and multiple scattering.  Otherwise
the positron energy and position resolutions would be degraded and the photon backgrounds
would be worse.
This requirement implied that no support structures for
the MWPCs could be located in the positron orbit region.  The positron spectrometer was
segmented in a way that minimized the occupancy in a high rate environment, especially in the
presence of magnetically trapped positrons.  The muon decay target had to have
sufficient mass to stop the muon, but conversely had to impose minimum mass through which 
the
positron would pass, which 
dictated a passive target.  These conditions were met by a target that was highly inclined with
respect to the beam.  This solution had the added
feature of spreading the stopping distribution along the beam direction, which allowed for
background suppression by requiring the photon and positron to originate from the same point
in the target.  

To achieve a sensitivity of 10$^{-13}$ in a reasonable time, the muon decay rate was set as high
as possible consistent with minimum confusion in positron tracking and acceptably low dead
times. 
The experiment was positioned in the LAMPF
stopped-muon channel
and set to accept the full muon beam, which was, at that time, the
most intense, high quality muon beam in the world.  The anticipated muon decay rate 
constrained the positron detector MWPC cell size to achieve acceptable occupancies.

The dominant $\mu \to e\gamma$  background comes from photons near 52.8 MeV 
in accidental time and spatial coincidence with a random positron emanating
from another muon decaying in the target.  Because the mass of
the positron MWPCs was very low, the most likely source of these photons was from the internal
bremsstrahlung (IB) process $\mu \to e\gamma \nu {\bar \nu}$.  Also high energy photons could
be generated by positron annihilation along its orbit.  The IB  process has a very low probability
of producing photons with energies close to
52.8 MeV, so good photon energy resolution provided a particularly important background
rejection
factor.  Furthermore, because the magnetic field confined charged particles to the central region,
the photon detector was expected to be very
``quiet" in the presence of a high muon decay rate.  This naturally led to an event trigger that
required a single high energy photon. 

Nonetheless, the trigger rate in the photon detector was expected to be $\sim$\,2
kHz, which would be too high to allow all of these events to be written to tape.   Therefore an
additional software
filter was needed to ascertain whether there was sufficient information 
in the photon spectrometer and then 
the positron spectrometer to support a possible $\mu \to e\gamma$ hypothesis.  This
on-line filter ran in a system of computers that were programmed with fast, reliable
code intended to make decisions either to keep or discard triggers.
The requirement on processing speed was driven by the need to analyze all triggers from a
LAMPF pulse and
write the selected candidate events to tape in the $\sim$\,8 ms period
between pulses.  The requirement of high reliability mandated extensive testing of the
on-line code with
a well-developed Monte Carlo (MC) simulation 
package prior to its implementation in the experiment.

The on-line filter was designed not to reject many $\mu \to e\gamma$ signal candidates,
and therefore, the
data written to tape would likely include large numbers of events that were not consistent with a
$\mu
\to e\gamma$ hypothesis.  Off line, 
the data were to be studied with a series of increasingly more rigorous, and more
time-consuming, analyses, each
reducing the volume of  data, and finally retaining only those events that were highly probable
$\mu
\to e\gamma$ candidates.
  
The accuracy of the experiment depended on regular and extensive temporal and spatial
calibrations to assure that the test of
time-coincidence and the requirement for precise spatial tracking were not systematically
compromised.   The temporal and spatial accuracy was affected by a wide range of
environmental parameters and apparatus characteristics, which needed to be
monitored in coordination with the
event information saved to tape.  Moreover, these calibration data required substantial
subsequent analyses to assure the accuracy and reliability of the $\mu \to e\gamma$ search.

\section{Experimental Details}
\label{sect:details}

\subsection{Overview}
\label{sect:expt-overview}

\begin{figure}[tbp]
 \begin{center}
     \epsfig{file=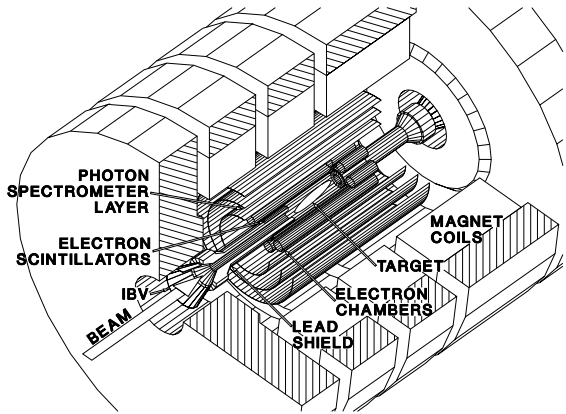,width=3.3in}
 \end{center}
\caption{A schematic view of the MEGA detector.}
%, showing: (1) the superconducting solenoid,
%(2) the incident muon beam, (3) a lead/heavimet annulus to stop the positrons, (4) the elliptical
%muon stopping target, (5) the positron spectrometer MWPCs, (6) the positron spectrometer
%scintillators, (7) the photon spectrometers, and (8) the internal bremsstrahlung veto detectors.}
\label{fig:3a1}
\end{figure}

Figure \ref{fig:3a1} shows a schematic view of the MEGA detector.  Both the positron and the
photon spectrometer systems were contained within the 1.5 T magnetic field produced by a
superconducting solenoid magnet.  A surface-muon beam from the LAMPF stopped-muon
channel entered the detector along the axis of the
superconducting solenoid.  The beam stopped in the elliptical target foil located at the center of
the detector.

Positrons from muon decay followed helical trajectories in the magnetic field.  They were
observed by an array of eight cylindrical,
high-rate
MWPCs that surrounded the stopping target.  The
MWPCs measured the crossing of positron tracks in all three dimensions, which allowed for the
determination of the muon decay point and the positron momentum.  Two annular
arrays of plastic scintillators, located near the ends of the positron wire chambers, provided
timing information.  After passing through the scintillators, the positrons entered thick
lead/heavimet annuli, where they stopped while producing a minimum of high-energy gamma
radiation.  Cylindrical plastic scintillators that were located inside the upstream and downstream
positron scintillator arrays and beyond the lead/heavimet annuli, ``ring counters", provided
timing calibrations.  The positron spectrometer had an outer radius of 30 cm, large enough to
contain all the positrons that were produced by muons decaying in the stopping target.

Photons from muon decay were observed in a set of three concentric, cylindrical pair
spectrometers that surrounded the positron spectrometer.  Each pair spectrometer utilized two
lead foils to convert high-energy photons into $e^+/e^-$ pairs.  The electrons and positrons were
then tracked through a set of drift chambers to determine the energy and propagation direction
of the original photons.  An MWPC located between the two convertor foils determined where a
given photon converted, and an array of plastic scintillators determined the conversion time.

The
hardware trigger for the experiment was designed to identify high-energy photons in the pair
spectrometers.  Events that passed the hardware
first- and second-stage triggers were read into a workstation where 
a partial analysis of each photon shower was performed, and then
the hits in the positron spectrometer were examined to determine if the minimum number
necessary to support a $\mu \to e\gamma$ hypothesis were present.  If so, 
the event was written to tape for
subsequent off-line analysis.

\subsection{Beam}
\label{sect:beam}

     The stopped-muon channel at LAMPF~\cite{thompson79} provided the muons
for the MEGA experiment using a 
surface-muon beam~\cite{reist78} tune.  The characteristics of
the beam were a flux of $2\:\times\:10^8$/s, 4\% muons above the kinematic 
endpoint from
stopped-pion decay (29.8 MeV/c) in the production target, and a ratio of 
positrons to
muons of 10:1.  To reduce the positron flux by a factor of 100, a 20.3-cm 
gap by 127-cm
long, crossed electric
and magnetic field separator, operated at a total voltage of 200 kV, 
was employed upstream of the last
focusing quadrupole.  The beam was tuned in the last lens to enter the 
solenoid with as
little loss as possible; the solenoid also had a strong focusing effect 
on the beam.
In order to produce an extended longitudinal beam-spot on the slanted 
target, the spot was purposefully
defocused to have a full width at half 
maximum (FWHM) of
3.5 cm (normal to the beam direction)
with the solenoid at nominal current.  When all beam tailoring with slits
was completed, the maximum stopping 
rate available
was $4\:\times\:10^7$ s$^{-1}$ for 1 mA of protons incident on a 6-cm
graphite production target.

The central momentum of the tune was 28.3 MeV/c
with a 10\% acceptance.  
A very small fraction of the muons
stopped in the
vacuum window between the beam line and the solenoid, 
and their decays were used for calibrating the upstream timing 
scintillators.  Following the vacuum window, a degrader foil 
was placed inside the heavy lead shielding upstream of 
the target.  The degrader thickness was chosen so that 
%in order to minimize the thickness 
%of the target,
%it was made to have a normal thickness of 76 $\mu$m.  
seventy-five percent of the muons
stopped in the slanted target.  The balance propagated to a 
foil that was inside the downstream shielding.  Essentially all of the remaining muons stopped
there, 
and their decays were used for calibrating the timing of the downstream
scintillators.  
%The final windows in the beam line are made of nearly 
%transparent Mylar
%to allow for detector alignment inside the magnet.
At this momentum, no pions survived to arrive at the experiment, and
residual positrons that passed the separator were focused by the solenoid and went
harmlessly through the apparatus.

     The procedure for calibrating the number of muon stops involved several steps.
Initially, with the magnetic field off, a beam of muons and positrons was 
brought into the center of the magnet.  The particles passed through a thin ion 
chamber, a 0.16-cm
thick scintillator 10 cm downstream, and a 1.3-cm thick scintillator another
10 cm downstream.  The separation between these elements 
helped to keep the solid angle for detecting
muon decay products low.
With low intensity beam, the number of muons were
counted in the first scintillator where they stopped, and 
the positrons were counted in the thicker scintillator.  The sum of the energy 
deposition
from both species was measured in the ion chamber.  The contribution of the 
muons to the ion chamber response was then deduced by placing a degrader
upstream of the ion chamber
to remove the muons.  Next, the field was turned on, with the ion chamber 
(electric field parallel to the magnetic field) upstream 
of a surface barrier Si detector, and the
ratio of muons to positrons was measured with the field on but at low rate.
Finally, with the field on, the ion chamber was placed
substantially upstream of a 
%500 $\mu$m 
target that stopped all the muons in the beam.  The count rate of several of the
upstream positron timing scintillators, which detected muon-decay products, was
compared to the ion chamber charge.  The ratios were observed to track linearly.  Thus a
calibration of the upstream positron 
scintillators to the muon-stop rate was made.  A comparison of the acceptance of
these scintillators to a MC simulation agreed to 5\% under the
assumption of 97\% muon polarization at the time of their decay.  The integrated
counts in the positron scintillators were then monitored throughout the data acquisition to
extract the total number of  stopped muons.

\subsection{Magnet}
\label{sect:magnet}

The MEGA detectors were located inside a large-bore superconducting solenoid
magnet.  This magnet was originally used in the Large Aperture Solenoidal
Spectrometer at SLAC \cite{LASSSLAC}.  For MEGA, only 3 of the original 4 superconducting 
coils were used and
the large opening in the downstream iron pole-piece was filled with iron except
for a small hole along the axis.  The solenoid had a clear bore diameter of
1.85 m and a clear bore length of 2.89 m.  The superconducting Cu-Nb coils were
immersed in a 4000 liter liquid Helium bath during operation; they carried 1178
amperes with a current density of $\sim$\,4000 A/cm$^2$, which produced a
1.5 T central field.   

Prior to inserting the spectrometers, the magnetic field was measured on a 5.1 cm
grid using Hall probes, which were calibrated in a uniform field 
measured with an NMR probe.  The principle component of the cylindrical
magnetic field was along the incoming beam direction, and varied
by $<$\,3\% within 1 meter of the magnet center along the central axis, with larger
variations at increasing distances off axis.  During data acquisition, the current in the
magnet was held constant to $<$\,0.1\%.  The detailed field map was tabulated for use in data
analysis.  

\subsection{Target}                               
\label{sect:target}

The muon stopping target \cite{koetke01} consisted of a planar sheet of 0.1 mm thick Mylar
oriented such that
the normal to the target plane was inclined at 83$^{\circ}$ with respect to the incident muon
beam.  This Mylar target was supported in space by rigid attachment to the inside walls of a
13 $\mu$m thick Mylar cylinder, 7.6 cm in diameter, giving the target a planar elliptical shape. 
The length of the target along its major axis was approximately 50 cm.
The supporting cylinder was filled with helium gas and was maintained in its rigid shape by a 20
torr differential helium gas pressure.  In this inflated form, the target was measured to be flat,
with deviations from flatness less than $\pm$\,1 mm.

The inclined thickness of the target as seen by the incoming beam was sufficient to stop 75\% of
the muons, passing only those that underwent significant multiple scattering.  By contrast, the
thickness
presented to outgoing positrons of interest was small, thereby introducing minimum energy loss,
multiple scattering and
annihilation.  The target inclination also spread the muon stopping distribution over a broad
range in $z$, enhancing the ability to distinguish between the separate origins of photons that
were produced in random coincidence with positrons.

Because the target was passive, the first point where the trajectory of a positron was measured
was
typically several centimeters away from the
muon-decay point.  Therefore, the vector momentum of the positron at the muon decay was
determined from the intersection of its observed helical track with the target plane.  This made
precise knowledge of the target location critical.  The position of the target when mounted in the
spectrometer was determined by direct visual measurements, based on a grid penned on the target
surface.  Approximately 100 points on the target were measured in all three coordinates using
survey instruments.  Their locations formed a plane with a fitted
$\chi^2$ of 1.1 for errors of 1 mm on the space points.  The absolute location of the target was
measured with the same precision when the target was in place for data acquisition.

\subsection{Positron Spectrometer}
\label{sect:positron-spect}

\begin{figure}[tbp]
 \begin{center}
     \epsfig{file=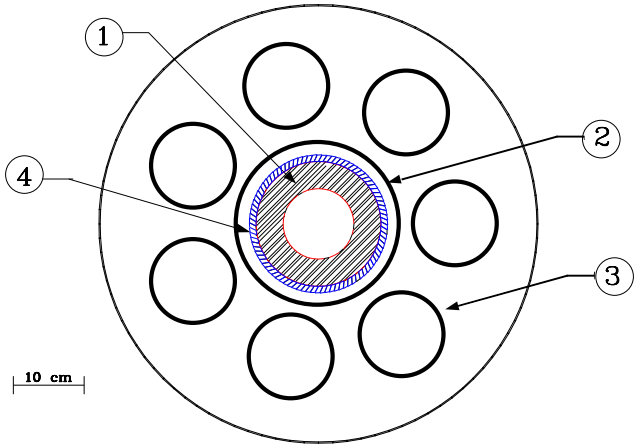,width=3.3in}
 \end{center}
\caption{\label{fig:3e1}  An axial view of the positron spectrometer showing: 
(1) the lead and
heavimet terminus for the positrons, (2) the Snow White MWPC, (3) the dwarf
MWPCs, and (4) the positron scintillators.  The outer circle represents the
tension shell.}
\end{figure}

The positron spectrometer \cite{cooper98} consisted of a large central cylindrical
MWPC
(called Snow White) and seven smaller cylindrical MWPCs (called dwarfs) as
shown in Fig.~\ref{fig:3e1}. The size, position, and internal design of these
chambers were chosen on the basis of MC studies to keep the occupancy
of the individual elements to $\sim$25\% with particle fluences of 4 $\times$
$10^{4}$ mm$^{-2}$\,s$^{-1},$ so as not to confuse the pattern recognition. 
(An original design of 3 concentric cylindrical MWPCs failed the occupancy
requirements.)
The overall diameter (60 cm) of the positron spectrometer was large enough to 
keep all orbiting positrons from entering the
photon spectrometer due to the 1.5 T solenoidal magnetic field.
The length of the chambers was chosen as 126 cm to match the solid angle
acceptance for the $\mu \to e\gamma$ signal within the photon spectrometer.

To minimize multiple scattering and to reduce the production of high-energy
photons, the chambers were built with an effective thickness of $3\times
10^{-4}$ radiation lengths, including the thickness of the cathode foils, the
gas in the chambers, and the wires. The design avoided structural supports in the MWPCs
anywhere in the detector volume where positrons might pass.  Accordingly, the wire tension was
maintained  by a cylindrical support, the ``tension shell", external to the positron orbit, and the
cylindrical shapes of the thin cathode foils were maintained by differential gas pressure.
The chamber anode wires were too long to be electro-mechanically stable under high voltage, so
low-mass garland supports were required to mechanically divide the wires into shorter regions.

\begin{figure}[tbp]
 \begin{center}
     \epsfig{file=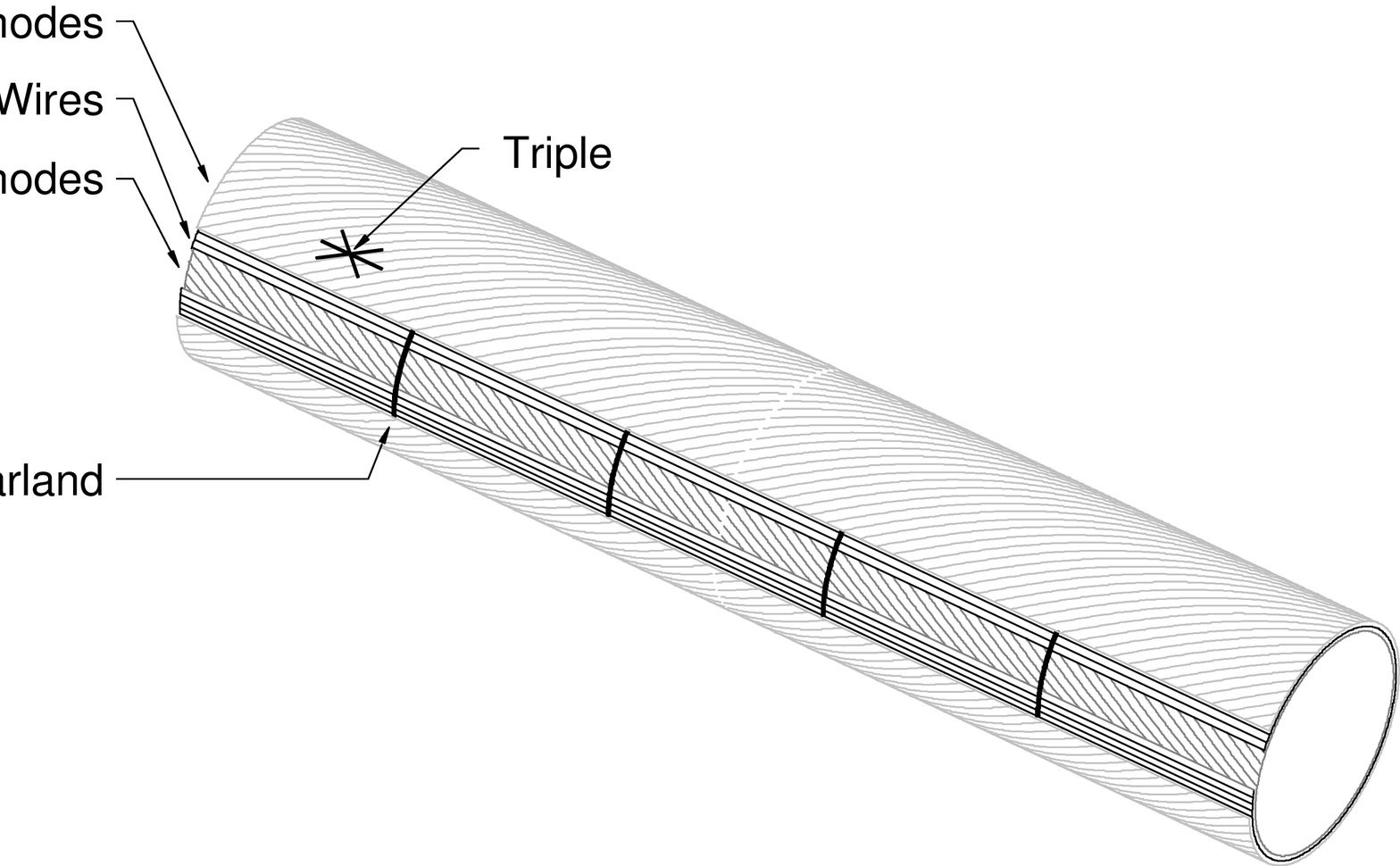,width=3.3in,clip=}
 \end{center}
\caption{\label{fig:3e2}  A partial cut-away view of a dwarf chamber showing 
the anode
wires between the two cathodes with counter-directed helical stripes.}
\end{figure}

\begin{table}[tbp]
\caption{Parameters of the MEGA positron spectrometer MWPCs.
\label{tab:3e1}}
\begin{tabular}{ll}
 Chamber length   & 126 cm \\ 
 Chamber radius   & 11.138 cm (Snow White)\\
                  & 5.982 cm (Dwarfs)\\ 
 Wire spacing     & 1.3 mm \\ 
 Wire type        & 15 $\mu$m Au-W \\ 
 Wire tension     & 25 g \\ 
 Half gap         & 1.778 mm \\ 
 Cathode foil     & Cu (200 nm) on Kapton (25 $\mu$m) \\ 
 Cathode stripe width & 2.7 mm (Snow White) \\
                      & 2.8 mm (Dwarfs) \\ 
 Cathode stereo angle & 29.05$^\circ$ (Snow White) \\
                      & 16.61$^\circ$ (Dwarfs) \\ 
 Chamber gas      & CF$_{4}$ (80\%) + C$_{4}$H$_{10}$ (20\%) \\
\end{tabular}
\end{table}

The basic layout of a dwarf chamber is shown in Fig.~\ref{fig:3e2}, and the
chamber
parameters are listed in Table \ref{tab:3e1}. The orientation of the anode
wires
parallel
to the magnetic field minimized the number of anode cells activated by a
helical positron path since the same cells would be struck multiple
times in the orbit. The spacing between anodes and the separation
between the anode high voltage plane and the cathode ground plane
were chosen to achieve the design resolutions at the decay point (2 mm in
space and 0.5\% in momentum) and to have acceptable occupancy rates.

A longitudinal position resolution of 4 mm for a track crossing was required to
obtain better than 0.5\% momentum resolution.  To achieve this goal,
the copper coating on the cathode foils was segmented into electrically separated,
helical shaped stripes, each of which had an individual readout.  
To reduce occupancy rates, the stripes were
separated into upstream and downstream channels by a division at the median
plane.  The copper
coating on the Snow White cathodes was divided into stripes only over the
scintillator region, leaving the copper coating in the central region unbroken
except for the division at the median plane.  
This prevented the high occupancy rate in the central region of Snow White from obscuring the
resolution required adjacent to the scintillators.  
%This cathode stripe geometry was also driven by the
%need to minimize occupancy rates.

A sophisticated gas system was built to mix the chamber gases and to maintain
the 20 torr
differential gas pressure that was required to support the chamber
cathode shapes.  In addition, the gas system was used to inflate the
muon stopping target support cylinder.  The high occupancy rate in the
spectrometer dictated that a fast chamber gas be used.  A freon/isobutane
mixture
of 80\% CF$_{4}$ and 20\% isobutane achieved this goal with an ion collection
time of $\sim$\,15 ns~\cite{HOBBIT}. Water vapor (0.2\%) was added to the gas
mixture to suppress continuous discharges. Also the inner and outer volumes
that
surrounded the MWPCs were filled with helium to reduce positron interactions
as they passed through the spectrometer.  

A LeCroy 1445 high voltage supply~\cite{LECROY} provided the high voltage
for the MWPCs.  It was linked via a serial connection to a computer running the
LabView~\cite{LABVIEW} instrument control program.  The LeCroy 1445 was
chosen because it provided an exceptionally fast trip response, which was
critical
for these chambers, while allowing computer-linked control and monitoring.  The
LabView controlling program was written to allow simultaneous independent
control of each of the 13 different high voltage supplies used. 

The positron chamber read-out system \cite{cooper98} was constrained by several requirements. 
First, the high instantaneous rate per wire ($10-20$ MHz) required a wide
bandwidth ($\sim$\,100 MHz) in order to resolve the chamber hits cleanly. 
Second, the high flux through the chambers made it necessary to run the MWPCs
at relatively low gas gain ($\sim$\,5 $\times$ 10$^4$), thus requiring highly
sensitive electronics with extremely low noise levels.  Finally, the limited
real estate available for the chamber-mounted preamplifiers and the large
number of channels restricted the design to a bare minimum of
components.  Preamplifier outputs were sent to rack-mounted
amplifier-discriminator cards within the experimental cave.  Discriminator
outputs were routed to Phillips 10C2~\cite{PHILLIPS} FASTBUS latches
for read out.

\begin{figure}[tbp]
 \begin{center}
     \epsfig{file=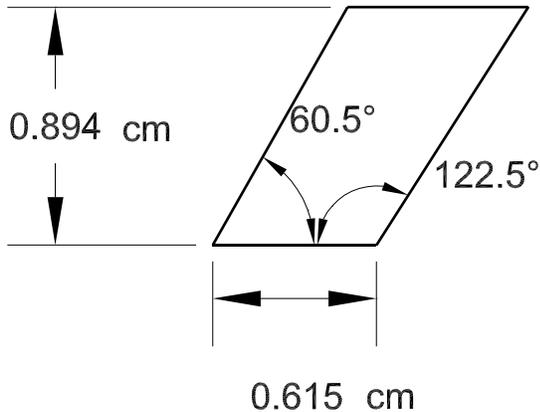,width=3.3in}
 \end{center}
\caption{\label{fig:4} Cross-sectional view of a positron scintillator.}
\end{figure}

Positron timing information was obtained from the 174 plastic scintillator strips 
that formed two cylindrical barrels.
Each scintillator was a 30-cm long rod with a trapezoidal cross section
(see Fig.~\ref{fig:4}).  This shape was chosen to minimize the number of scintillators
crossed by a single spiraling positron.  Each scintillator was individually wrapped in
aluminized Mylar foil for optical isolation.  The end opposite the
light guide connection was blackened to prevent multiple reflections.
The scintillators were
closely packed into a barrel on the outside of the lead/hevimet 
absorber.
One scintillator was missing to allow space for a tube to supply gas to the space 
between
the target and Snow White.  Each scintillator was connected to a phototube
with a 1.8-m long optical fiber light guide so the phototubes could be located
in a shielded enclosure outside the magnet.  Signals from the phototubes went
to FASTBUS ADCs and to discriminators, which fed
FASTBUS TDCs~\cite{PHILLIPS} and a logic OR for trigger purposes. 

An unexpected property of the detector, combined with the response of the electronics,
limited the rate capability of the experiment. As noted above,
the anode wires were placed parallel
to the magnetic field to minimize
their occupancy. A similar, but less easy to
understand, degeneracy existed for the helical cathodes.  If the same
read-out
channel was hit repeatedly within the resolving time of the electronics (20
ns), the pulses piled up.  This pile-up greatly increased the dynamic range
requirement on the electronics.
The channel-to-channel cross-talk isolation of the preamplifier
cards was measured to be 30 dB.  However, if a great
deal of energy was deposited near the same wire by several crossings of a
positron, a very large pulse occurred that induced cross talk on neighboring
wires.  Such an event had a small probability, but at high muon
stopping rate, this small probability was multiplied by the large number of channels in the entire
system so that there was cross talk in nearly every trigger.
To
reduce the effect of the cross talk, both the muon stopping rate and
the gain of
the chambers were reduced below design values.   The result was a limit on the instantaneous
muon stop rate of $2.5\times 10^8$  Hz, and average anode and cathode
efficiencies of 95\% and 85\%, respectively.

\subsection{Photon Spectrometer}
\label{sect:photon-spect}

The photon detector \cite{barakat,gagli88,chen,barber01}
consisted of
three independent, concentric, cylindrical pair spectrometers, which
surrounded the positron spectrometer.  Pair spectrometers were
chosen for detection of the decay photons because they provided some
directional information, and comparatively, the best possible combination 
of energy, timing, and spatial resolutions for the 52.8 MeV
photons of interest.  Increased conversion efficiency, while maintaining good
energy resolution, was achieved by using a system of multiple
spectrometers.  The detector had fine granularity, which
allowed high data rates with low event pile up.

\begin{figure}[tbp]
 \begin{center}
     \epsfig{file=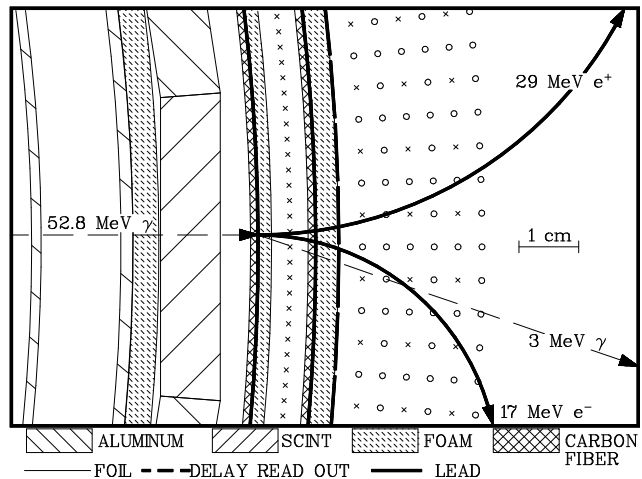,width=3.3in}
 \end{center}
  \caption{A cross section of a pair spectrometer layer, showing the aluminum
 support cylinder for an inner layer, and the timing scintillators,
 conversion cylinders, MWPC and drift detectors for the next outer
 layer. A typical conversion in the first conversion cylinder is shown.
 \label{fig:3f1}} 
\end{figure}

A cross sectional view of a pair spectrometer sector is shown in 
Fig.\@ \ref{fig:3f1}, and Table \ref{tab:3f1} gives radial dimensions of the
components.
Each pair spectrometer was about 175 cm in axial length.  It contained
two cylindrical, lead convertor foils between which was 
a MWPC.  Just outside the converter cylinders was a set of cylindrical
drift chambers to track the conversion pairs.  A scintillation barrel, placed
inside the conversion layers, timed the 
traversal of the then back-traveling lepton
pairs, and also measured their trajectory diameters, which were proportional
to their transverse momenta.  This information was used in the
trigger to select only the conversion of high-energy photons.  The
scintillators were also used to determine the time of conversion and to
establish the time reference for the drift chamber TDCs.

The tension of the chamber wires 
in each layer was balanced by the compressional resistance of 
a rolled and welded aluminum cylinder, 2 mm in thickness, which was placed
beyond the maximum turning radius of reconstructable $e^{+}e^{-}$ tracks.
This cylinder also acted as the outer gas barrier for the drift chambers.

\begin{table}[tbp]
\caption{Components of the photon pair
spectrometers.  All radii are in cm.
\label{tab:3f1}}
\begin{tabular}{lrrr}
 COMPONENT & Layer 1 & Layer 2 & Layer 3 \\
\hline
\multicolumn{4}{l}{Scintillator}\\
\multicolumn{1}{l}{~~Number of scintillators}&40&60&80\\
\multicolumn{1}{l}{~~Radius}& 32.0&47.9&63.4\\
\hline
\multicolumn{4}{l}{Lead conversion cylinder}\\
\multicolumn{1}{l}{~~Radius of 1st}&33.6&49.5&64.9\\
\multicolumn{1}{l}{~~Radius of 2nd}&34.5&50.4&65.8\\
\hline
\multicolumn{4}{l}{MWPC}\\
\multicolumn{1}{l}{~~Number of wires}&416&640&832\\
\multicolumn{1}{l}{~~Radius}&34.2&50.1&65.5\\
\hline
\multicolumn{4}{l}{Drift chambers}\\
\multicolumn{1}{l}{~~Number of wires}&208&320&416\\
\multicolumn{1}{l}{~~DC1 radius}&35.3&51.3&66.7\\
\multicolumn{1}{l}{~~DC2 radius}&36.1&52.1&67.5\\
\multicolumn{1}{l}{~~DC3 radius}&36.9&52.9&68.3\\
\multicolumn{1}{l}{~~DC4 radius}& & &71.1\\
\hline
\multicolumn{4}{l}{Aluminum support cylinder}\\
\multicolumn{1}{l}{~~Radius}&47.6&62.8&88.3\\
\end{tabular}
\end{table}

Each of the three spectrometer layers contained 3 conventional, 
cylindrical drift chambers with
individual drift
cells approximately 1 cm long and 0.8 cm high.  A fourth set of 
drift chambers was added to the outer layer.  These cells, except for
the inner drift chamber of each layer, were defined
by sense wires at the center of the cell, and field wires positioned
at the corners and between each sense wire.  In the case of the inner
drift chamber, the outer surface of the
conversion cylinder formed an equipotential on which a delay line \cite{barber01} was
positioned under each sense wire.

A typical event had a pair conversion in either the inner or outer lead
conversion foil, each being 0.045 radiation lengths.  
If the conversion occurred on the inner foil, the
pair traversed the MWPC and the outer foil layer before entering the
drift chamber region.  Thus, a signal in the MWPC at the vertex of an
$e^{+}e^{-}$ pair identified these specific events.  The $e^{+}e^{-}$
pair curled in opposite directions through the drift chamber region; their
trajectories were used to determine the transverse component of their
momenta.  After making circular arcs through the drift chamber
region, the pairs then passed back through the conversion cylinders, the MWPC,
and the scintillator. Depending on energy loss and multiple scattering in the
scintillators and the aluminum support cylinder, the pairs could continue
to spiral in the magnetic field, although with smaller radii, so that
in cross section the trajectories did not overlap.  Tracking
information was utilized only from the first pass of the leptons
through the drift chambers, although multiple
scintillator hits were used to improve event timing.

The $z$ location of the pair vertex
and the position where each lepton
spiraled
back through the conversion cylinder were measured with
the delay lines.  These points determined the polar (dip) angle of the
photon with respect to the cylinder axis.  The dip angle was combined with the measurement of 
radius of curvature of the leptons in the magnetic field to compute the full energy of the photon
and
its propagation direction.  The position of a drift-cell crossing
was determined with a precision of about 200 $\mu$m in the direction transverse to the magnetic
field and about 1
cm in the $z$ direction.  The time of the photon conversion was obtained 
by correcting the time-appearance of the 
scintillator signals for the flight time of the leptons.                             

The delay lines were
composed of  an etched laminate of 34 $\mu$m thick copper
sandwiching a 75 $\mu$m thick polyamide foil.
%Characteristics of these lines are given in Table \ref{tab:3f2}.
The signals were read differentially at each end by a voltage 
preamplifier feeding a constant fraction discriminator, and the
resulting signals  were then gated into a FASTBUS TDC to 
record the timing information from each delay line.
Each line was thus equivalent to a distributed parameter
transmission line, and because the resistance was not 
negligible compared to the 
characteristic impedance of the line, the signal rise time degraded and the 
pulse amplitude decreased as a function of distance along the line.
Although constant 
fraction discriminators were used to reduce the effect of the time walk
of the signal, this effect was, nonetheless,  a significant problem, and
limited the spatial resolution to about $\delta L/L$ = 0.5\% for the
approximately 175 cm long lines. 

\subsection{Internal Bremsstrahlung Veto}
\label{sect:ibv}

      The dominant source of photon background was the IB
process.  High energy photons emitted in IB have a 
high probability to be accompanied by a 
low energy positron.  In contrast the probability of low energy positrons from normal 
muon decay is small.  The IB 
positrons were easily and cleanly detected in coincidence with high energy 
photons in the Crystal Box detector \cite{bolton88}.  
The MEGA detector was instrumented 
with internal bremsstrahlung veto scintillators (IBV)
mounted on the surface of the pole tip penetrations, 
whose purpose was to veto low-energy positrons emitted in 
the IB process.  
In the 1.5-T magnetic field, 0 to 
5-MeV positrons were constrained to follow the field lines and thus 
hit the 30-cm radius penetrations in the magnet iron surrounding the 
beam.  The time of arrival of these positrons ranged from 20-80 ns after the 
muon decay, as determined by the initial dip angle.  During the 
development of the beam for the experiment, it was discovered that beam 
halo prevented IBV counters from being effective upstream.
Consequently four counters were mounted in the
top, bottom, left, and right positions that were only used 
to monitor the beam position.  As no muon beam reached the 
downstream IBV location, it was lined with 18 counters, arranged azimuthally, 
to intersect all positrons in this energy range originating in the 
target with a positive value of $p_z$.

\subsection{Trigger and Data Acquisition}
\label{sect:trigger}

The first-stage trigger \cite{chen} used the fact that the sum of the diameters of
the circular trajectories of the conversion pair measured the transverse
component of the photon momentum.  MC studies of photon yields were
used to select hit patterns in the MWPC and scintillators of the photon spectrometers for
coding into the first-stage trigger electronics.  Each photon pair spectrometer operated
independently producing its own first-stage trigger.
The trigger required that at least two scintillators were hit in spatial coincidence with three
groups of MWPC cells, corresponding to a minimum width consistent with a 
$\sim$\,37 MeV/c transverse photon momentum ($p_T$).
Thus, the trigger cutoff was determined by $p_T$ 
and not total photon energy.  However,
$p_T$ was strongly correlated with total photon energy
because of the geometry of the detector.

        The purpose of the second-stage trigger was to reject events that satisfied the
first-stage trigger, but did not produce tracks in the drift chambers, or did not have
the sufficient number and pattern of hits in the drift chambers to reconstruct a
photon event.  It imposed additional constraints on a
candidate photon event based on hit requirements in the drift chambers,
to reduce substantially first-stage triggers caused by
accidental coincidences between two lower-energy photons, or true high-energy
photons produced by conversions in either the photon scintillators or the aluminum
cylinders.  Each photon spectrometer layer
had its own second-stage trigger module.  Table \ref{tab:3g1}
compares various trigger rates at full beam intensity.

\begin{table}[tbp]
\caption{A comparison of MEGA trigger rates at 
full beam intensity (in Hz).
\label{tab:3g1}}
\begin{tabular}{cccc}
Item&Per $\mu$ Stop&Instantaneous&Average\\
\hline
%\medskip
$\mu$ Stop&$1.$&$2.6 \times 10^8$&$1.3 \times 10^7$\\
$1^{st}$-stage trigger&$1.2 \times 10^{-4}$&$3.1 \times 10^4$&$1.6 
\times 10^3$\\
$2^{nd}$-stage trigger&$6.4 \times 10^{-5}$&$1.7 \times 10^4$&$8.3 
\times 10^2$\\
Photons $>$\,47 MeV\tablenotemark[1]&$8.5 \times 10^{-6}$&$2.2 \times 10^3$&$1.1 
\times 10^2$\\
\end{tabular}
\tablenotetext[1]{Reconstructed in off-line analysis from data with the
on-line filter
for the positron spectrometer temporarily disabled.}
\end{table}

An electronic controller, the
``routing box", was used to coordinate the triggers from all three layers
with the  dead time of the data acquisition system, and 
was the central control module for the trigger system.  It
combined the first-stage trigger results with selected additional signals to form a desired set of
triggers, then routed the gate outputs to the group of FASTBUS modules
that were required to read out the event.  It also kept track of the
FASTBUS-module busy signals, sent start signals to the second-stage trigger,
monitored the result of the second-stage trigger decision, and sent
out a fast clear or transfer-to-memory signal (i.e., keep the event for
subsequent read out) to the appropriate FASTBUS modules
following the second-stage trigger decision.  All of these tasks
were accomplished under computer control.

Each pair spectrometer had dedicated FASTBUS
TDCs, ADCs, and latches~\cite{PHILLIPS} that could be read out independently to
minimize the dead time.  However, the
positron spectrometer needed to be read out for every event, regardless of which
photon layer generated the trigger.  To facilitate this, the TDCs and ADCs for each positron
scintillator were read by alternating between a pair of modules, 
designated ``even" and ``odd".  This allowed the data
acquisition system to accept, for example, a trigger in layer 1 while a
previous trigger from layer 2 was still being digitized.
In contrast, the transfer-to-memory
cycle of the FASTBUS latches was sufficiently
fast that only a single set of read-out modules was needed for the
positron spectrometer MWPCs.
The typical data acquisition system dead time was $\sim$\,6\%.

Since the event rate in the photon pair spectrometers was low, the number of FASTBUS
interface modules was reduced by multiplexing the photon chamber signals.  A
scheme gating a block of these signals in the azimuthal location of
the first-stage trigger minimized accidental overlap for the wire chamber signals \cite{chen}.

The first-stage trigger rate as filtered by the second-stage trigger was still too
high to commit all event data to permanent storage.  Therefore events were entered
into an eight-computer DECStation 5200 workstation farm that further
filtered their number via preliminary physics analysis. 
Each event that contained a candidate photon
with an energy $E_{\gamma} > 42$ MeV passed the on-line photon spectrometer
requirement.
Typically 27\% of the triggered events passed the photon on-line pattern
recognition requirements, and these events were then analyzed by the positron
spectrometer on-line code ARC described in Sect.\@ \ref{sect:arc}.  If the event also
satisfied the ARC criteria, it was written to tape for subsequent off-line filtering. 
Typically, 15\%
of the events that passed the photon on-line requirements also
passed ARC.  In addition, a small sample (0.5\%) of events that failed the on-line
event filter was written to tape in order to provide a continuous monitor of the
performance of the on-line filter codes.

\subsection{Data Taking Procedures}

Data for this experiment were accumulated during calendar years 1993-95.
Approximately every eight hours, the scintillator timing and
electronic pedestal calibrations were checked.  Otherwise, data were collected at the full 
operating rate except for the brief interruptions required for
the supplementary measurements described in Sects. V and VIII.  

During routine data collection, the performance of the detector was
monitored in a number of ways to insure the quality of the data.  Approximately 
80 separate measurements, such as room temperature, wire chamber voltages
and currents, solenoid field setting, cryogenic instrumentation readouts, etc., were
recorded roughly once per minute by a dedicated computer interfaced by
GPIB bus to sensors.  Quantities outside preset limits caused an alarm to sound.
Another system monitored quantities that were scaled and translated into quantities
such as dead time, pass-through rates in the software filters, detector rates, etc.
In the case of changes in accelerator parameters, the experimenters adjusted
the average stopping rate by changing the channel slits to keep the instantaneous rate
at 250 MHz.  In addition, a graphical single
event display (SED) was available for viewing the patterns of hits in the detector
elements on an event-by-event basis.

Deciding that an instantaneous stop rate of 250 MHz
was optimal proved to be quite challenging.  This decision was 
made after visually studying thousands of events
from the SED. As the final positron
spectrometer reconstruction
program was not available at the time the data were taken, the
complexity of events that could be properly analyzed had to be judged
{\it a priori}.

\section{Monte Carlo}
\label{sect:monte-carlo}

The MC simulation of the MEGA apparatus was based on the
EGS4 \cite{egs4} package, with several modifications to meet the requirements
of this experiment.  The geometries of the positron and photon spectrometers
were coded into the simulation according to their actual construction.  The
detector was subdivided into single-medium regions that were bounded by
up to seven surfaces (planes, cylinders, or cones).  Checking geometric limits
is recognized to be one of the most time-consuming tasks in any particle
physics simulation, and much effort was expended to optimize this
task.  For example, separate sections of the code were exercised
to find the intersection(s) of either a straight line segment or an arc of a
helix for each type of surface.  As a result the performance of the simulation was about four
times faster than would have been achieved with a GEANT-based \cite{geant3}
simulation.  Other special features of the simulation are presented below.

\subsection{Event Generation}

Events were generated within the simulation program using one of the
following processes:  (1) unpolarized $\mu^+ \rightarrow e^+\gamma$,
(2) polarized $\mu^+ \rightarrow e^+\nu\overline{\nu}$,
(3) unpolarized $\mu^+ \rightarrow e^+\gamma\nu\overline{\nu}$,
(4) $\pi^{\circ} \rightarrow \gamma\gamma$ (where the $\pi^{\circ}$
originated from $\pi^- p \rightarrow \pi^{\circ} n$),
(5) $\pi^- p \rightarrow \gamma n$,
(6) $e^+$ with uniformly distributed momentum and direction,
(7) $\mu^{\pm}$ with a cosmic ray energy and direction spectrum.

For the first three processes, the muon was assumed to be at rest in the thin,
planar, Mylar target.  For the fourth and fifth processes, the $\pi^- p$ reaction
was assumed to occur with the $\pi^-$ at rest within a thick cylindrical
polyethylene target.  Finally, the simulated cosmic ray spectrum had a $1/E^{2}$
energy dependence and a $\cos^2\zeta$, zenith angle dependence.

For the muon decay modes, an option was provided to simulate the high-rate
backgrounds.  In the positron spectrometer, this could be done by superimposing hits
from either MC simulated muon-decay positrons or real events 
onto the high-rate data.  In the photon spectrometer, the overlay was done by including
additional random hits according to the relative frequencies that various patterns
were observed during background studies.  These options permitted a reliable
determination of the detection efficiency as a function of beam rate.

\subsection{Extensions to EGS4 Physics}

A routine extracted from GEANT was added to the EGS4 package to permit the
tracking of muons, including beam muons and energetic cosmic rays, within the
apparatus. (Cosmic rays were used in detector alignment and position resolution
studies, for example.)  Muons were treated as long-lived ionizing particles; neither
decays nor nuclear interactions were considered.

In addition to the limitations on the step size of charged particles
imposed by EGS4 and the geometry of the MEGA apparatus, the ionization
energy loss per step was not allowed to exceed 5\% in the positron spectrometer or
2\% in the photon spectrometer.  

For the traversal of charged particles through thin media -- gases or very thin
solids -- the mean number of elastic scatterings $N$ often falls below the value of
20.  Thus the Moli\`ere parametrization of the scattering angle
is no longer valid.  To account properly for single and plural scattering
in such cases, the Moli\`ere parametrization was replaced with a new
algorithm, if the mean number of scatterings was below 25.  In this new
algorithm, the number of scatterings in one step was sampled from a
Poisson distribution with mean $N$.  Then this number of single scatterings were
sampled from a screened Rutherford scattering distribution and convolved to
give the overall scattering angle.  The resulting distribution merged
smoothly into the Moli\`ere approximation for $N \ge 25$.

The {\it average} restricted energy loss by charged particles in EGS4 was replaced
with a new algorithm that incorporated fluctuations.  This was required to model
accurately the energy resolution of the spiraling positrons in the positron
spectrometer, and the fluctuations in pulse height detected by the MWPCs.  If the
Moli\`ere number of scatterings in one step was smaller than 25 or if the charged
particle was passing through the chamber gas of an
MWPC, the restricted energy loss was calculated using the method of
Talman \cite{talman} to sample the energy loss due to resonant as well as
non-resonant ionization of electrons from the various atomic shells of
the element(s) of the media.  (Resonant ionization results in an electron-ion
pair where the liberated atomic electron has no kinetic energy; non-resonant
ionization results in an electron-ion pair where the liberated electron
has kinetic energy below the EGS4 discrete-tracking cutoff of 25 keV.)  As
a side benefit, the individual sampled ionizations within an MWPC cell
were used to generate the detected pulse -- including fluctuations in pulse
height and arrival time at the anode.  If the Moli\`ere number of scatterings
exceeded 25 -- i.e., when traversing thick media -- the restricted energy loss
was sampled from a set of eight tabulations of the Vavilov distribution.
These tabulations were used to simulate energy loss in the photon
spectrometers, independent of the thickness of the media.

\subsection{Signal Generation}

Signals from energy deposited in the detector elements were recorded in simulated ADCs, TDCs
and latches exactly matched to the data format of the actual apparatus.
Additional simulation-specific information keyed to these signals was written to
help identify in detail the history of each event and debug the 
reconstruction and analysis programs.

The signal from energy deposited in each positron scintillator was propagated, with
a time delay, to the phototube and then recorded in an ADC and
TDC.  TDC dead times were simulated by imposing a time window from a
previous
over-threshold hit so that the second of two close-together hits was lost.  Each
TDC had a programmable pulse-height threshold
that was
tuned to match the detection efficiency of the actual scintillator.   Each ADC
had a threshold (for zero suppression) to match the behavior of the real ADCs.

Electrons released by energy deposition in each positron spectrometer MWPC cell were
propagated to the anode
wire.  The cell boundaries were tilted from radial by about $17^{\circ}$ due
to the Lorentz $E\times B$ effect on the drifting charge in the
chamber gas.  The arrival time of the avalanche -- usually
but not always from the point of closest approach of the trajectory 
to the anode -- determined the initial time of the anode and cathode signals.
The pulse on the nearby inner and outer cathodes was the spatial image of the
anode pulse with a gaussian shape and an rms size determined by the chamber half-gap and the
cathode stereo angle.  These initial pulses on the anode and cathodes were
then propagated with delay to the readout end of each element and turned into an electronic
pulse shape with a base width of about 25 ns.  For each event, the
electronic pulses on all channels were superimposed to account for possible
pulse pile-up due to  recurring hits on the same channel from multiple loops, as well
as uncorrelated hits at high beam rates.  The resulting signal was
discriminated using a threshold that was matched  
channel-by-channel, then checked to see
that it was within the time gate that had been started by the event trigger.
Signals that were above threshold and in time  were recorded in simulated latches.

In the photon spectrometer, the signal from energy deposition in a scintillator was propagated to
each
end and recorded in simulated TDCs and ADCs.  Drift chamber, delay line and
MWPC hits were simulated by determining the earliest arrival of an electron
cluster at the preamplifier mounted on the end of a sense wire or delay line.  
At the end of the event, the outputs of the ADC
for each scintillation channel and the TDC for each scintillation or wire chamber
channel that was above threshold were smeared to account for both
pulse-height independent and pulse-height dependent resolution effects.
Background noise hits in the scintillators, MWPCs, delay lines, and drift
chambers were superimposed on the event.  The event was then examined to
determine if the hits were sufficient to pass the hardware first- and
second-stage trigger requirements.  If so, the event was saved after adjusting all
TDC times so that the hardware
first-stage trigger defined the effective zero time.  (During the simulation, $t=0$
was given by the muon decay time.)

\section{Detector Calibrations}

\subsection{Photon Detector Drift Times}
\label{sect:drift-calib}                                         

The magnetic field significantly influenced the drift time in the wire
chambers.  Given the approximate electric and magnetic fields and the
equilibrium drift velocity, the Lorentz angle for the drifting electrons was
approximately 37$^\circ$.  This large drift angle increased the drift time by about
a factor of 2, and reduced the ionization charge collected on the wires.  In
addition, some of this charge leaked between adjacent drift cells for certain track
geometries.  Nevertheless, MC simulations demonstrated that the equal-time
drift contours were approximately circular, although charge deposited in drift-cell
corners remained trapped for long periods \cite{barakat}.
These simulations also indicated a
slight angular asymmetry of the contours. This effect was ignored in the analysis
of the drift-distance vs.\@ drift-time data for all three wire chambers of all three
layers.  The drift time distributions were fit by one quadratic function:
\begin{equation}
d = v_{0}\,T + a_{0}\,T^{2}.
\end{equation}
The parameters of the fit, $v_{0}$ = 0.0049 cm/ns and $a_{0}$ = 
$-9 \times 10^{-6}$ cm/ns$^{2}$, were very close to those obtained from the
MC simulation, averaged over all wire chambers and layers.  By this
procedure
the distance from the drift wire to the tangent of 
the track position was located within 0.2 mm rms, which was more
than sufficient for the experiment, as resolutions were dominated
entirely by the axial, rather than the azimuthal, position measurement.

\subsection{Photon Wire Chamber and Scintillator Efficiencies}

Representative wire chamber efficiencies (magnetic field on) are given in Table
\ref{tab:3i1}, as measured by cosmic rays tracked through the system.  The
uncertainties in these numbers are about 1\%.  The table shows that the wire
chamber efficiencies were relatively stable from the beginning, 1993, till 
the end, 1995, of the experiment.  Scintillator efficiencies were
somewhat lower than expected, and Layer 3 in particular developed
several dead channels,
%after an accident when the photon pair spectrometers
%were dropped from a crane at a height of $\sim$\,30 cm.
resulting in its lower efficiency.  Layer 3
was added to the experiment
after the cosmic ray data were taken in 1993, so
there were no 1993 data for comparison.

\begin{table}[tbp]
\caption{Wire chamber and scintillator efficiencies by layer and by year.
\label{tab:3i1}}
\begin{tabular}{lccccc}
\multicolumn{1}{c}{Wire Chamber} & \multicolumn{5}{c}{Efficiencies} \\
\hline
         & \multicolumn{2}{c}{Layer 1} & \multicolumn{2}{c}{Layer 2} &
\multicolumn{1}{c}{Layer 3}\\
\hline
 Year & 1995 & 1993 & 1995 & 1993 & 1995 \\
\hline
MWPC & 98.9 & 99.1 & 100. & 99.1 & 97.5 \\
DC1 & 97.1 & 96.8 & 97.0 & 97.6 & 96.4 \\
DC2 & 97.5 & 98.8 & 99.5 & 95.9 & 95.6 \\
DC3 & 97.2 & 98.5 & 96.2 & 95.9 & 96.1 \\
Scintillators & 95.1 & 96.5 & 96.8 & 97.5 & 87.6 \\
\end{tabular}
\end{table}

\subsection{Spectrometer Alignment}

It was necessary to determine the location of the positron spectrometer elements relative to the
established
MEGA coordinate system in order to reconstruct the positron tracks precisely.  The spectrometer
coordinate system was defined to have its $z$-axis
aligned with the cylindrical axis of Snow White. 
The positive direction pointed downstream along the muon beam.  The $x$-axis was defined
to lie along the centers of Snow White and the MWPC, dwarf number one, which
was approximately horizontal.  The $y$-axis was directed upward completing a
right-handed coordinate system.  The axis of Snow White was assumed to
lie
along the axis of the solenoidal magnetic field, and no evidence of 
misalignment was found.  The (0,0,0) location was defined to be on the Snow
White axis at the center of the symmetric
upstream-downstream helical cathodes.

The spatial alignment of the spectrometer components was accomplished using
cosmic ray tracks, with and without the magnetic field, and positron
helical tracks from muon decays in the target \cite{cooper98}.
The various relative orientation parameters were adjusted to optimize the
$\chi^2$ fit to a large ensemble of such tracks.  The process was sequential,
and began with an alignment of the axial anode wires in the seven dwarf
MWPCs relative to the anode wires in Snow White.  This alignment required only
fits to the cosmic ray tracks as viewed along the $z$-axis, and 
resulted in the $x$-$y$ position of the centers of each of the dwarf
chambers.  It also determined the azimuthal rotation angle of the anode wires around the
cylinder axis for each dwarf chamber.  Once the full set of  21 parameters was determined, the
axial ($z$) alignment was begun.  A track
crossing the MWPC produced an anode cluster and an inner and an outer 
cathode cluster.  By rotating the inner and/or outer cathode cylinder, the
intersection in $z$ along the two cathodes  was brought into agreement.  This 
determined the relative alignment of the cathodes
in the eight MWPCs.  The inter-cathode
alignment between the dwarf chambers and Snow White was established by
``sliding" each dwarf cathode inner and outer pair along the $z$-axis to optimize
the track fits. 

In all, 36 alignment parameters were determined for the positron
spectrometer.  It was initially assumed that
the anode wire spacing around the cylinder axes was sufficiently
regular that wire-by-wire adjustments would not be required.  During the
alignment process, however, it was discovered that, while the spacing
between neighboring wires was always well within the 75 $\mu$m chamber
winding tolerance, there were regions where many
sequential wire spacings were systematically large or small.  The cumulative irregularities led to
misalignments of some MWPC wires by a significant fraction of a cell.
Therefore, wire-by-wire azimuthal position tables were
constructed for the anode wires \cite{cooper98} and employed to position the wires properly in
the positron spectrometer MWPCs.

The alignment of the photon spectrometer relative to the positron spectrometer
was determined using helical cosmic ray tracks.  Photon scintillators were
used as a trigger and coincident wire chamber information in the photon and
positron spectrometers was read.  For the azimuthal alignment, circle
segments, seen when viewed along the $z$-axis, were reconstructed using wire
chamber hits from the positron and the photon spectrometers.  The track fits
were optimized by varying azimuthal rotation offsets and the ($x$,$y$) location of
the cylinder axes for the wire chambers in each of the three layers of the
photon spectrometer. The rotational corrections for the three photon
spectrometer layers were found to be $<$ 15 mr, and in good agreement
with optical survey results.  In
addition, a test was made for misalignment of the cylinder axis of the
photon spectrometer layers with respect to the $z$-axis of the coordinate system;
the corrections were found to be negligible.  All of the alignment
parameters were included in the detector database and employed in the track
reconstruction programs.

\subsection{Delay Line Calibrations}

Delay lines were used to determine the axial position of events in
the pair spectrometers.  Since the relation between position along the lines
and propagation time was observed to be very
linear, the axial position was determined by the propagation velocity of
the signal, ``slope'', and the time-zero offset, ``intercept'', for
each line.  These constants were extracted by analysis of 
cosmic tracks through the complete MEGA detector.  With the field
on, cosmic rays passed through the pair spectrometers and positron
chambers, and 3-dimensional event positions extracted from the
positron chambers were used to 
project the trajectory arcs through the pair spectrometers.  As the
position in each pair spectrometer was determined from the innermost
layer outward, the calibration constants for each delay line and drift
chamber were obtained.  

It was possible for ionization along a chamber track to cross
cell boundaries and produce signals on adjacent delay lines,
but events were also
observed that appeared to be induced on adjacent lines by
electronic cross talk.  
These events were
removed for the calibration analysis.  
The propagation velocity along the line was 0.654 cm/ns for
single-line signals and 0.634 cm/ns for
double-line signals \cite{barber01}.

\subsection{Positron Spectrometer MWPC Efficiencies}

A measurement of the positron spectrometer MWPC performance was obtained
using cosmic ray trajectories \cite{cooper98}.
The trajectories were reconstructed from
the photon spectrometer wire chamber hits and then projected through the
positron MWPCs. 
These tracks were used to measure the individual efficiency of
every anode and cathode in each of the eight MWPCs.  Plots of these
efficiencies were used to reveal inoperative electronics channels,
misaligned connectors, and overall MWPC performance. For data analysis these
efficiencies were superseded by high
statistics efficiency studies carried out with positron tracks from low
rate stopped muon decays in the target, as described in Sect.\@
\ref{sect:low-rate-michels}.

From the individual anode and cathode efficiencies, the average anode and
cathode efficiency in each MWPC was computed at various anode high voltage
settings.  Plots of these average efficiencies were used to establish the
optimum operating voltages and demonstrate that the MWPCs were on
plateau.  A further detailed examination of the individual channel
efficiencies as a function of high voltage gave evidence for
non-uniformities in the shape of two of the MWPCs.  These difficulties
were addressed by using multiple high voltages applied to different
regions of these two MWPCs.

The average anode and cathode efficiencies for each MWPC were measured as
a function of the delay between the photon spectrometer trigger and the
gate applied to the positron MWPCs, thus establishing the correct gate
timing in the trigger.  In addition, average anode and cathode
efficiencies were employed to set the optimum value for the thresholds on
the amplifier-discriminator cards for the MWPCs.

\subsection{On-Line Timing}
\label{sect:on-line-timing}

The positron-photon time difference, $t_{e\gamma}$, was calculated in
terms of time intervals that were measured by the apparatus,
so that a value of zero corresponded to
coincident emission from the target.  Part of this calculation involved
the subtraction of the invariant time intervals due to fixed electronic
delays or signal propagation in detector elements and cables.
These equal-time offsets were measured in special timing calibration runs
that determined the offset for each scintillation channel, and 
monitored its drift as the local environment changed.
Separate equal-time offsets were recorded for the even and the odd
read-out TDCs on each positron scintillator.

\begin{figure}[tbp]
 \begin{center}
     \mbox{\epsfxsize=3.3in \epsfbox{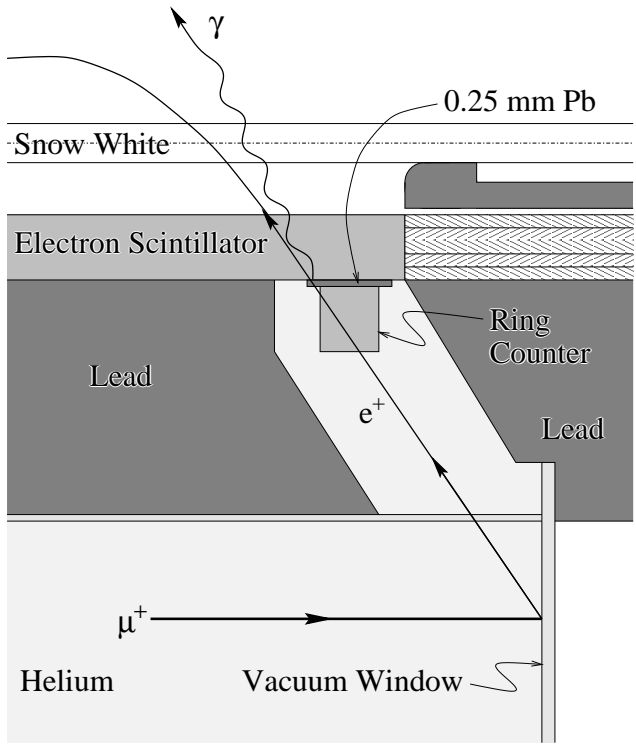}}
%     \mbox{\epsfxsize=3.3in \epsfbox{fig_s3_i_3_new.eps}}
 \end{center}
\caption{The location of the downstream Ring Counter.}
\label{fig:3i3}
\end{figure}

The on-line time calibration used a pair of dedicated ring-shaped scintillators
located at the far ends of the upstream and downstream beam pipes and
inside the positron scintillator barrels (see Fig.\@ \ref{fig:3i3}). Each
counter was instrumented to record the hit times at both azimuthal ends of
the ring (``high-$\phi$'' and ``low-$\phi$'').  A small slot in
the lead beam pipe permitted some positrons produced in muon decays on the
vacuum window to hit the ring counter. Some of these positrons then
radiated a
photon; the bremsstrahlung probability was enhanced by wrapping the
outside of
each ring counter with lead tape. The positron nearly always hit a
positron scintillator, while the radiated photon entered the MEGA detector
roughly
along the slot, then converted
into an ${e^+}/{e^-}$ pair or Compton scattered in the photon spectrometer.

The scintillator timing calibration events were triggered by the coincidence
of a ring counter and any photon scintillator.  In this restricted
geometry, the typical flight paths of the photon and positron
were fixed on average so that the centroids of the
time distributions were constants for the struck scintillators.  
%(The
%calculated values for the photon flight path
%for layers 1, 2, and 3 were $c \cdot 0.9\,\rm ns$,
%$c \cdot 1.5\,\rm ns$, and $c \cdot 2.1\,\rm ns$, respectively.)  
Offsets
were determined channel-by-channel relative to the ring counters.
Global offsets between the upstream and downstream ring counters and
between the even and odd positron scintillator TDCs were determined from
averages over the photon scintillators.

Timing calibration runs were taken once every eight hours during the course
of the data runs, and new timing constants derived from these runs were loaded
into the on-line event filter programs once
per day.  This ensured that the timing constants in use by the on-line
filter at any given time were accurate to $\lesssim$\,1 ns.

\section{Photon Event Analysis}
\label{sect:photon-recon}

\subsection{Event Reconstruction}

The photon spectrometer reconstruction algorithms needed to find $e^{+} e^{-}$ 
pairs resulting from the conversion of a photon, and to determine the photon
energy, propagation direction,
conversion point, and conversion time from the tracks produced by the 
pair.  When viewed along 
the magnetic field direction, the $e^+$ and $e^-$ traveled 
in circular orbits, activating drift cells along their paths. 
As the pair passed through detector material and lost 
energy, the radii of the orbits decreased. These characteristics defined the 
patterns of cells that were hit in a typical photon conversion. 
The transverse momenta of the electron and positron were
determined by fitting the hit drift 
cells to circular paths, and the longitudinal momenta
of the particles were calculated from 
the $z$ values at the vertex and edges of the event as measured by the delay
lines. The time 
of the photon conversion was determined from the times 
of scintillators that were hit in each circular orbit of the particles. 

Pattern recognition was an important consideration during the 
design of the photon spectrometer hardware, so parameters such as MWPC wire 
spacing and drift cell sizes were determined from consideration of mechanical and
physical tolerances, ability to reconstruct events and system cost.  Consequently,
many photon events were straightforward to analyze.  The
tracks from the $e^+ e^-$ pair formed
clean circles of decreasing radii in a view of the detector perpendicular to the drift
wires.  The occupancy rate in the drift chambers, coupled with the active gating
scheme mentioned in Sect.\@ \ref{sect:trigger},
produced relatively clean events.  The challenge was
to develop reconstruction algorithms to
automate what the eye could see, in spite of
the wide statistical variations from event to event.  

The reconstruction process passed through several stages of computer code. 
Initially candidate events were found by looking for patterns of hit cell 
coordinates that were similar to typical event patterns found in MC 
simulations. The cells corresponding to the best guess for 
those hit during the initial arcs of the conversion pair were then tagged. Since 
each member of the pair passed through a substantial amount of material, 
subsequent arcs were only useful for helping to improve the timing resolution 
for the photon conversion.  Equal time contours were found for each hit cell 
as described in Sect.\@ \ref{sect:drift-calib}. The drift time information 
for these cells was then used in a 
non-linear least-squares fitting routine to determine the best circle fit for each 
member of the pair. Constraints on the circles were included to ensure that the 
pair originated at a common location. By incorporating the 
$z$-information for the tracks, the location,
 ($R_{\gamma}$,$\phi_{\gamma}$,$z_{\gamma}$), and
time, $T_{\gamma}$, of the photon
conversion and the vector momentum
of the photon were determined.  Appendix
\ref{sect:photon-anal-algorithm} gives a description 
of the analysis routines.

\subsection{Background}

Background events arose from a number of sources including 
reconstruction errors, two low energy photons that converted close together in 
space and time and Compton scattering of a photon. The pattern recognition 
algorithms eliminated most ($\approx$\,90\%) of the Compton events. 
Most of the events where two lower energy photons converted and were 
reconstructed as a single high energy candidate were eliminated by 
additional checks on the number of cells that were hit outside of the 
reconstructed photon shower.  However, some candidate events survived due to 
these background sources, which produced a high energy tail on the 
reconstructed photon energy spectrum. The final step in the reconstruction 
process was a visual examination (``hand scanning") of candidate events. 
The sample chosen for hand 
scanning included all events that passed the initial $\mu \to e\gamma$ 
cuts imposed in the off-line filter. Interspersed with this sample were MC 
events, which were used to determine the efficiency of the hand scanning 
process. Most background events that survived the filter cuts were removed 
from the final event sample during the hand scanning process. 

\subsection{Performance}

The overall efficiency of the photon spectrometer to detect 52.8 MeV gamma rays
was 2.4\%.  This number was obtained by estimating the efficiency with 
a high-statistics MC simulation, then correcting that estimate for event
losses during hand scanning and effects that were not simulated in the MC.
The hand-scanning efficiency was measured to be 0.91, as described in 
Sect. \@ \ref{sect:offline-filt}.  
The primary effects that were not simulated properly by the MC were electronic
cross talk and charge migration from one drift cell to its neighbors.  Initially, 
the correction for MC deficiencies
was estimated to be 0.85 from a study of many real and MC events with the 
single-event display.  Subsequently, this correction was measured to be 0.84 during
the $\pi^0$-decay studies (Sect.\@ \ref{sect:pi0}).

The overall efficiency of the photon spectrometer may be understood as follows. The ideal upper
limit for the efficiency was 5.5\%. This was determined by simulating 52.8 MeV photons from
$\mu \to e\gamma$ decays with the MC assuming all detector channels were perfectly
efficient, selecting those events that passed the hardware first- and second-stage triggers, then
analyzing them using the actual space points from the $e^+$ and $e^-$ trajectories recorded in
the MC history files. This estimate neglected event losses due to finite detector
efficiencies and realistic pattern recognition, in addition to those mentioned above. The
correction for finite detector efficiency was 0.91. It included comparable contributions due to the
scintillators and the delay-line drift chambers, and smaller contributions due to the MWPCs and
the other drift chambers.

The expected correction for pattern recognition losses was estimated to be 0.70. 25\% of the
events contained additional hits in the vicinity of the vertex or an edge from subsequent $e^+$ or
$e^-$ loops that made it impossible to identify the initial $e^+$ and $e^-$ tracks uniquely. In
practice, approximately half of these events reconstructed properly nonetheless. However, the
remaining events often had fitted energies well above the true photon energies. Therefore, 
all such
events were rejected to obtain a significant reduction in background at the expense of a modest
loss of efficiency. 7\% of the events consisted of $e^+ e^-$ pairs with highly asymmetric energy
sharing that could be reconstructed using the MC space points, but not from the drift
chamber information alone. For these events, the track of the low energy particle included four
space points in the MC history file, but one or two of the corresponding drift distance
measurements were obscured by hits from the high energy particle on the same cell. Combining
these effects, the expected efficiency of the photon spectrometer was 2.7\%. The 10\% deviation
from the true efficiency was caused by the remaining inadequacies in the pattern recognition
and reconstruction codes.

\section{Positron Event Reconstruction}

\subsection{On-line Pattern Recognition}
\label{sect:arc}

An analytic reconstruction code (ARC) was used in the
on-line filter to determine the search region in the positron MWPCs and
scintillators for hits that supported a $\mu \to e\gamma$ hypothesis.
The code used the coordinates, ($R_{\gamma}$,$\phi_{\gamma}$,$z_{\gamma}$), and
time, $T_{\gamma}$, from the on-line photon pattern
recognition code to examine a
limited range of positron scintillators. 
These scintillator ranges were established by systematic MC studies of $\mu \to
e\gamma$ candidate events and set to include all positron candidates for a given
($R_{\gamma}$,$\phi_{\gamma}$,$z_{\gamma}$).  If no hit positron scintillators were found
within 32 ns following the photon trigger, the event was abandoned. 
Adjacent positron scintillators within the range with TDC times within
2 ns were assumed to be associated with the same positron.

For a selected hit scintillator in the range, a triple coincidence in Snow White, consisting of a hit
anode
and an overlapping pair of stereo cathodes, was demanded within an azimuthal window
adjacent to the hit scintillator.  This triple coincidence indicated the passage of a positron
through Snow White into the scintillator.  If no
triple coincidence was present, the scintillator was not used and the next scintillator in the
range was examined.
If a triple coincidence was found, it was assumed to provide a space point
($x_{sw}$,$y_{sw}$,$z_{sw}$) along the $\mu \to e\gamma$ positron orbit, and the scintillator
provided the orbit end-time, $T_e$, from calibrated TDC information.  (The time $T_e$ was
corrected for light propagation along the scintillator.)                

The photon and positron information were used to obtain an analytical estimate of the path of a
$\mu \to e\gamma$ positron through the positron spectrometer. (The details of this orbit
calculation can be found in Appendix \ref{app:arc}.)  In the calculation, the positron was {\it
assumed} to originate from a $\mu \to e\gamma$ decay at rest in the slanted target. The positron
emerged with a momentum {\bf p}$_e$ (52.8 MeV/c) directed opposite to the momentum of the
trigger photon {\bf p}$_{\gamma}$ (52.8 MeV/c). The estimated positron helical track was then
projected through its intersections with the positron MWPCs where anode-cathode triple
coincidences were expected. Using MC generated $\mu \to e\gamma$ decays, windows were
established around the calculated positron trajectory. The size of these windows was tuned to
optimize the acceptance for the $\mu \to e\gamma$ signal while rejecting photon triggers with no
$\mu \to e\gamma$ positron present (background).

If the number of missing triple coincidences in the windows along the positron track did not
exceed an established maximum, the event was written to tape as a $\mu \to e\gamma$ candidate. 
If, however, the number of missed triple coincidences exceeded this limit, the hit scintillator  was
not used and the process was repeated for the next hit scintillator in the 
possible range. If the event
was not kept after all hit scintillators in the range were examined, the entire process was
repeated with the value of $T_e$ adjusted by
$\pm$\,2 ns to allow for possible on-line timing misalignments. During the 1994 and 1995 run
periods, the value of $T_e$ was also adjusted by $\pm$\,4 ns if necessary, and the event checks
repeated. The increased search provided a better measure of the background time spectrum in the
vicinity of the
$\mu \to e\gamma$ signal region.

The performance of the ARC filter was benchmarked on MC signal events,
generated assuming design chamber efficiencies, overlaid either on real
data or on simulated backgrounds. With this overlay, the success rate could be measured at the
full intensity of the beam. The filter found 89\% of the $\mu \to e\gamma$ events within the
nominal geometric acceptance, while passing no more than 0.1\% of the background.  The
selection reduced the number of taped events by approximately two orders
of magnitude.  This pass rate was much less than the 15\% quoted in Sect.\@ \ref{sect:trigger}
because less stringent criteria (e.g., broader windows, fewer hits) were ultimately 
required to define an event as acceptable in the on-line filter.  The 11\% loss in 
signal acceptance was due in large part to losses associated with
anode and cathode inefficiencies.
An additional small loss was attributed
to failures in the iterative calculation of the orbit in cases where the positron orbit was
nearly parallel to the planar target at the muon decay point.

\subsection{Off-line Event Reconstruction}
\label{sect:positron-offline}                

Two closely related reconstruction codes were employed in the off-line analysis
of the positron data \cite{stantz97}.  The first applied when the rates were low, and attempted to 
find all tracks in the event.  This approach was not feasible in the high rate
environment during data acquisition.  Thus the second algorithm 
was confined to a search in the phase space where a candidate track
had some reasonable probability of being a $\mu \rightarrow e \gamma$ event.  The two
programs had many features in common, but the special
treatment of high rates is noted below.  Unlike the photon reconstruction program, this code was
challenged to find real events that at high rates were not easily visualized.  

The logic of the reconstruction was to define the clusters of electronically struck wires
induced by the positrons as they crossed the chambers, and these clusters were
grouped into potential space points.  Disregarding the non-uniformities in the
magnetic field, a good track swept out a helix as the positron propagated from its
decay point to the scintillators.  The space
points defined above projected to a circle when observed in an end view,
and formed a straight line when unrolled in a coordinate system
consisting of the longitudinal position and the turning angle measured from the decay 
point.  Hence, the algorithm looked for circles amongst the candidate space points, and
used these circles to search for the full three-dimensional track.  At high rates, tracks
made from uncorrelated hits, ``ghost'' tracks, were reasonably probable, so that 
stringent quality criteria were required to ensure, with good probability, that  
tracks were real.  
A summary of the techniques used to construct the space points and, 
ultimately, the tracks is given in Appendix \ref{app:pos-pattern}. 

The track reconstruction algorithm described in Appendix \ref{app:pos-pattern}
ignored the impact of energy loss
and multiple scattering in the fitting process.  Hence, the resolution 
was not optimal and the computed energy was the average of the positron energy at the two
ends of the track.
A final least-squares fit was done to include these effects. Hits from a
track were fit to calculated trajectories propagated through the detector system.  Good tracks
were required to have a ${\chi}^2$ per degree
of freedom below 3.0.  
The parameters were the position and momentum components of the positron at the first
chamber crossing after leaving the target.  Later, the final fit track was stepped back to
the target to get the properties of the decay.
During propagation, the mean energy loss in each material
was subtracted from the energy of the particle.  Hence, at the decay point, the fit was a
good approximation to the decay energy.  Multiple scattering was taken into account
in the design matrix \cite{dataanal} (the inverse of the weight matrix).  
Contributions to the position resolution of the predicted hits were found numerically
and added in quadrature with the intrinsic resolution of the chamber hits.  As a given
anode wire could be hit several times by a positron, the resolution of these wires was
summed over all the loops.  Thus a single anode wire was fit only once and had a
single weight for all loops.

\subsection{Performance}                

\begin{figure}[tbp]
   \begin{center}
     \mbox{\epsfxsize=3.3in \epsfbox{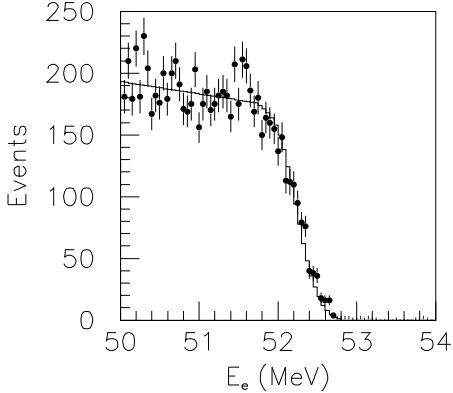}}
   \end{center}
   \caption{The $E_{e}$ spectrum from $\mu^{+} \rightarrow e^{+} \nu 
\overline{\nu}$
extracted from low-rate data.  The solid 
curve is the
fit used to extract the energy and resolution of the edge.}
   \label{fig:l-r-m}
\end{figure}

The best measure of the performance of the track reconstruction and fitting algorithms
is the positron energy spectrum described in Sect.\@ \ref{sect:low-rate-michels}.
The three features to note are the absolute energy of the kinematic threshold, the
width of the edge, and the existence of unphysical events above the cutoff. A typical 
low-rate muon-decay spectrum is shown in Fig.\@ \ref{fig:l-r-m}.  The central 
energy of the edge is 52.21 MeV, the resolution is 188 keV rms, and
there is very little high-energy tail.  The value of the central energy is about 0.6 MeV below 
$m_{\mu}$/2 because the fit momenta were not corrected for energy loss.  The resolution is 
much closer to the MC simulation, which predicts 161 keV rms, 
than the high-rate data because the space
points are less confused with the overlapping hits in the chambers; the 
residual difference between
MC and data is due to cross talk in the electronics.  

\begin{figure}[tbp]
   \begin{center}
     \mbox{\epsfxsize=3.3in \epsfbox{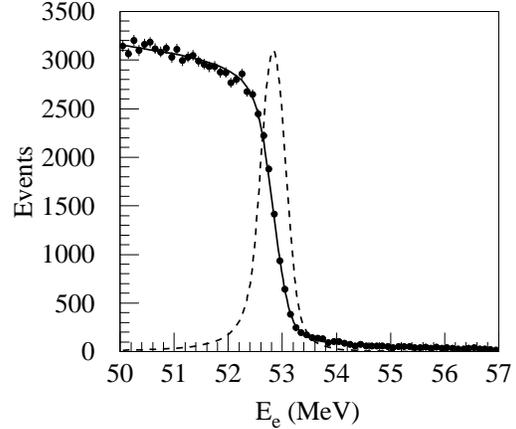}}
   \end{center}
   \caption{The $E_{e}$ spectrum from $\mu^{+} \rightarrow e^{+} \nu 
\overline{\nu}$
extracted from full rate data. 
%for the middle topology group.  
The solid curve is the fit used to extract the line shape (dashed curve).}
   \label{fig:5b2}
\end{figure}

The positron energy spectrum observed at full rate
is displayed in Fig.\@ \ref{fig:5b2}.  The absolute
energy is within 10 keV of $m_{\mu}$/2, indicating that the mean energy loss
was properly calculated in the final fit.  The energy resolution is 230 keV $\sigma$
at 52.83 MeV.  The 
resolution must be compared to the MC simulation of the positron line shape
from the $\mu \rightarrow e \gamma$ process.  For low rate data this value is 
180 keV for events with the same ``topology''; i.e., the same number of loops and 
chambers traversed.  When these simulated events were overlayed onto a background of
high rate events, the resolution degraded to 210 keV.  This value is in reasonable
agreement with data, given the inadequacies of the simulation with respect to noise
in the detector.  About 3\% of the events are above the kinematic cutoff.  These
events are unphysical, and are composed of fragments of different tracks that 
fooled the reconstruction algorithm.  They cause a minor reduction in the efficiency
for detecting $\mu \rightarrow e \gamma$.

\begin{figure}
  \begin{center}
    \mbox{\epsfxsize=3.3in \epsfbox{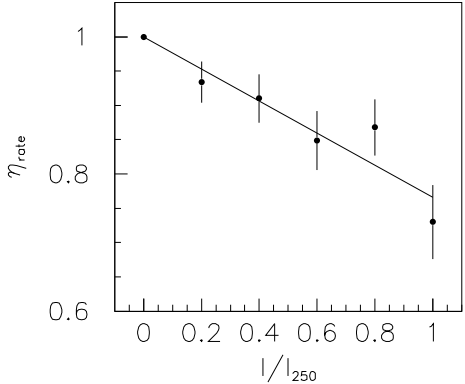}}
  \end{center}
  \caption{The rate dependence of the efficiency of the reconstruction 
code.  The relative efficiency, $\eta_{\rm rate}$, is defined to have
a value of 1 at low muon-stopping rates.  The relative rate is defined as
the ratio of the instantaneous beam intensity, $I$, to an instantaneous intensity
of 250 MHz, $I_{250}$.}
  \label{fig:5b3}
\end{figure}

The efficiency of the track finder was rate dependent.  As discussed in
Sect.\@ \ref{sect:low-rate-michels}, the low-rate efficiency of the track finder 
was 65\%.  As the rate increased to the operational intensity of 250 MHz (instantaneous),
another 23.5\% of the
events were lost.  The linear dependence of the efficiency of the track finder on rate is 
illustrated in Fig.\@ \ref{fig:5b3}.  This plot was created by taking a set of
reconstructed positron tracks from normal muon decay, selecting those above 50
MeV, and overlaying them onto real background events of varying rates.  If the same
events were found with the same energy and position properties to within 2$\sigma$,
the events were considered to be properly reconstructed in the high rate environment. 
A similar result was found by overlaying simulated events on the real background.

\section{Supplementary Measurements}
\label{sect:calibdata}

A number of supplementary measurements were performed during the experiment.  Several
utilized
dedicated data sets that were taken under special conditions in order to test the energy, timing
and tracking resolutions of the positron and photon spectrometers, and the ability of the MC
codes to simulate these resolutions reliably.  In addition, various subsets of the $\mu \to
e\gamma$ production data events were used to improve the time resolution of the spectrometers
and to optimize the global timing offsets between the positron and photon spectrometers.

\subsection{Normal Muon-Decay Studies}
\label{sect:low-rate-michels}

Data were taken at a low instantaneous
muon stopping rate ($\sim$\,500 KHz) using a
positron scintillator as a singles
trigger to validate the positron MC and 
to determine the discriminator threshold for each anode
and cathode channel in the MWPC electronics.  The thresholds were deduced from
channel-by-channel measurements of the chamber efficiencies and tabulated for
input to the MC code.

To determine chamber efficiencies, the low-rate version of the reconstruction
code was used to fit tracks to all available hits except at the
crossing studied.  The anode wire efficiencies were measured using
0-loop events (the loop number denotes the number of complete circles that a positron 
made through the spectrometer before hitting a scintillator), because multiple-loop events
generated additional ionization,
which increased the efficiency.  The cathode efficiencies were determined from
1-loop, multiple-dwarf events, which were the least contaminated.
The optimum size of the window, which was used to determine the presence of a
hit, was 7(5) anodes (cathode strips) wide. 

The measured efficiency was averaged over the length of each anode wire or
cathode strip.  Typically there were significant variations along the length
of a wire or strip.  For example, the efficiency was very low in the vicinity
of the garlands.  Each anode wire that was not connected to the high
voltage created a ``dead spot" on each
cathode it crossed.  Also, non-uniform gaps (due to deformed cathode foils
\cite{cooper98})
created different gains in some parts of the chambers.  These factors made the
measured efficiency sensitive to the population of tracks along the wire or
strip.

The measured efficiencies were used to generate an initial set of thresholds
for the MC. Then a set of events were generated and the
channel-by-channel efficiencies were computed for these events. When the
efficiencies from the MC were compared to those from the data, there
were significant differences due primarily to differences in the populations
of events.  However, from the differences, a correction to the thresholds
could be calculated and a new set of MC events was generated.  The
modified thresholds altered the population of events to bring them into 
closer agreement with 
events reconstructed from the data.  After a few iterations
(usually three) the agreement between data and MC was better than 1\%
for most regions of the chambers.

Several other comparisons between the low-rate data and MC confirmed
that the MC correctly simulated the geometric and kinematic
acceptance of the detector.  These included matching the energy distribution of
positrons for 0, 1, 2, and 3 loop events, and the loop-number distributions
for all events and for events with positron energies above 45 MeV.

Data taken at low rates with a 0.25 mm thick vertical target were used to
calibrate the absolute energy scale for positrons.  This was accomplished by
fitting the edge of the muon-decay spectrum separately for groups of events that
made 0, 1, or 2 loops and were traveling either upstream or downstream.  The six values for the
edge energy were compared to values from the MC, including the
effects of energy loss in the target and detectors.  Then a scale factor of 1.0096 was applied
to the magnetic field map to match the data to MC; this achieved a
precision of 10 keV in the absolute energy scale.

\subsection{``Hole Target" Data}

A good test of the positron pattern-recognition algorithm was the analysis of
data taken with a ``hole target".  An elliptical target with a 2 cm $\times$
10 cm rectangular hole 
replaced the normal target positioned at the center of the MEGA
detector.  Reconstructed muon-decay positron tracks measured the known edges
of the hole and determined the fraction of (incorrectly) reconstructed tracks
that appeared inside the target hole.

\begin{figure}
  \begin{center}
     \mbox{\epsfxsize=3.3in \epsfbox{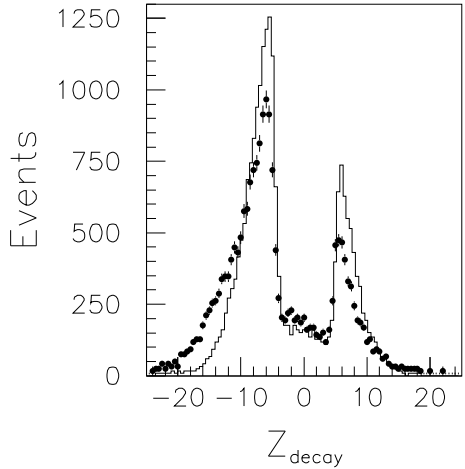}}
  \end{center}
\caption{Target intersection coordinates of low-rate muon-decay events when
the coordinate in the perpendicular direction fell in the range of the hole in
the target.  The points represent measured events, and the curve 
represents a MC simulation.}
\label{fig:6b1}
\end{figure}

Figure \ref{fig:6b1} shows a comparison of the observed intersections of data
and MC positron helices with the target plane in a reference frame aligned
with the plane of the target. The edges of the hole are clearly evident in the
projection and the agreement between data and MC confirms that the position
resolution is properly simulated.  The differences in the overall shape of the
projections are due to the use of a Gaussian shaped beam for the MC that
did not match the shape of the actual beam.

Many of the events that reconstructed within the target hole intersected the
target at a very shallow angle and thus had poor position resolution in the
plane of the target.  Also, some events intersected the target at
two points during their first loop.  For these events, both intersections were
kept even if one fell in the hole. The total fraction of events lying inside
the hole was 4.9\% in the data and and 3.3\% in the MC simulation. The
difference resulted mostly from MC inadequacies in modeling the positron
chamber efficiencies and detector noise.  The discrepancy was larger for
multi-loop events, which had a higher probability of passing near a garland.
Likewise, detector noise produced extra chamber hits that degraded the spatial
resolution and caused pattern-recognition failures.

\subsection{Off-Line $e-\gamma$ Timing}
\label{sect:offline-timing}

\subsubsection{Timing Within the Photon Spectrometer}

The photon scintillators served the dual purpose of providing fast timing
information to the trigger, and photon timing relative to the positron.
Dual-threshold discriminator circuits were used,
instead of constant fraction discriminators, since the scintillators were subject
to multiple hits from conversion pairs.  The discriminators
produced timing signals that had a pulse-height dependent slewing; to
obtain the optimum timing resolution, it was necessary to correct for this.

On-line calibration of the photon scintillators relative to the ring counters
was discussed in Sect.\@ \ref{sect:on-line-timing}.
Calibration constants were updated continuously 
during the data acquisition.  However, on-line calibration did not include
any correction for pulse height.  Also, background in the histograms 
degraded the resolution.  As part
of the off-line analysis, codes were developed to correct for pulse-height
slewing and to improve the time resolution using MEGA data.  
In Sect.\@ \ref{sect:photon-recon}, the photon reconstruction algorithms
were discussed. 
Timing corrections were performed using scintillator information obtained
from the first-pass scintillators for events where both members of the
conversion pair had well isolated
3-side edges.  During the
filter process, timing information for reconstructed photons satisfying this
criterion was written to an output data stream and then processed with a
separate timing analysis code.  

The first-crossing arrival time of the
$e^+$ and $e^-$ at their scintillators should be the same when
corrected for the flight time from the vertex.  This flight
time was calculated in the off-line filter code as part of the information
determined from the circle fits.  This time difference was not the same if
pulse-height slewing caused a shift in one or both outputs, or if timing jitter
appeared due to background problems. 
Data from successive calibration runs were used to
obtain correction factors for these two effects.  

The procedure used to obtain the timing correction factors began by determining
the apparent time difference between the two edges of the events.
Fits to the time difference spectra were used to
produce an $N_{scint}$ $\times$ $N_{scint}$ matrix of time differences
for each layer.  In
order to determine the coefficients needed to minimize the time jitter, the
matrix was inverted subject to the constraint that the
off-line timing correction factors averaged to zero
within each photon layer.
Using these corrections, the data were replayed
to obtain, event-by-event,  the time at each end of a scintillator after correcting for
the time-of-flight along the scintillator.  A fit was made to this arrival time
versus pulse height for each photomultiplier tube to obtain a pulse-height
slewing correction.  A final pass through the data allowed the pulse-height
correction constants to be applied to the time differences, and then the time
offset corrections were recalculated.  Following this procedure the photon
scintillator timing was improved by a factor of $\sim$\,1.8  relative to the on-line timing
resolution.  These pulse-height slewing and alignment timing constants were
then used in the final event analysis. 

\subsubsection{Timing Within the Positron Spectrometer}

The positron spectrometer scintillators also needed to be
timed with respect to each other and corrected for pulse-height dependent slewing.
As with the photon spectrometer, the scintillators
in the positron spectrometer were timed with respect to each other by
calculating time differences between scintillators that should be in time, plotting these time
differences
versus the pulse-height in the scintillator, and forcing the average time difference
to be zero.  The same mathematical methods were used in both spectrometers, 
but the data selection was somewhat
different.  In the positron spectrometer, data from positron
scintillator clusters were collected, and
the time differences between adjacent scintillators, the charges on the
scintillators, the $z$ position of the hit on the scintillator, and the
information of whether the cluster came from an even or odd event
were stored.  The $z$ positions of the
scintillator hits were used to correct the scintillator times for the
transit time in the scintillator.  Slewing corrections and time offsets were
then calculated for all upstream and downstream phototubes, for both even
and odd events.

\subsubsection{Timing Between the Photon and Positron Detectors}
\label{sect:rc-heg-timing}

Section \ref{sect:on-line-timing} describes the routine timing calibrations
that were performed
over the course of the experiment, and the previous subsections
describe the improvements
to these timing constants that were obtained during the off-line data
analysis.  However, while
the off-line timing constants significantly improved the overall time
resolution, they were derived
``locally", within a given photon layer or group of positron scintillators, so
they did not ensure
that the timing offsets were optimized between the two spectrometers.  This
match was achieved by
studying production data events that contained coincident hits in a ring
counter and a positron
scintillator, together with a high-energy photon shower.

During the off-line event filtering (see Sect.\@ \ref{sect:offline-filt}),
those events that contained
a fully reconstructed high-energy photon and a ring counter hit were
selected for further analysis
if the photon propagation direction projected back to the vicinity of the
ring counter and the time
difference between the high-energy photon and the ring counter hit
satisfied a loose coincidence constraint. 

These events were then subjected to several additional cuts to
isolate those with coincidences between a positron scintillator cluster (one or more
contiguous positron scintillators in time coincidence with each other) and the
high-energy photon shower.  These cuts
included: (1) the $z$ location of the photon conversion must be
consistent with the ring counter and photon layer that were hit; (2) the positron scintillator
cluster must be in time coincidence
with the ring counter; (3) the $\phi$ coordinate of the positron scintillator
cluster must be consistent with both the $\phi$ location of the high-energy
photon shower and the $\phi$ location
of the ring counter hit.  Also there must be no more than one additional
positron scintillator cluster within a $\pm$\,15 ns time window and
a $\pm$\,1 rad angular window in $\phi$ about the time and location
of the high-energy photon shower on the same end of the positron spectrometer as the hit in the
ring counter.  Furthermore, if a second
positron scintillator cluster was present, it must be out of time with
respect to the ring counter hit.

\begin{figure}[tbp]
 \begin{center}
     \mbox{\epsfxsize=3.3in \epsfbox{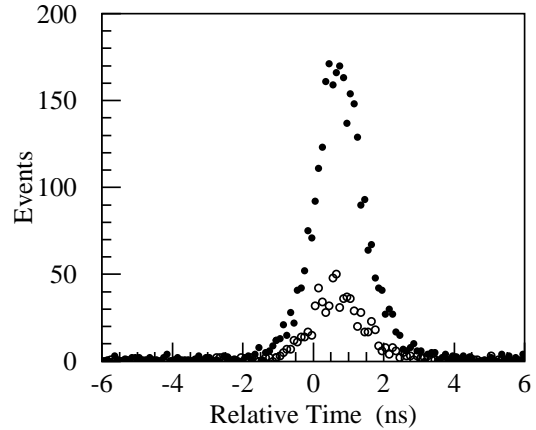}}
 \end{center}
\caption{Time difference spectra between positron
scintillator clusters from downstream, even TDCs and
high-energy photon showers in layer 2 for events that contained
coincident ring counter
hits.  The open circles show
the distribution for a group of 204 runs that were taken early in
1994, before the on-line
filter code was modified to enrich the sample of ring counter, high-energy
photon events.  The solid circles show
the distribution for a group of 211 runs that were taken later
in 1994, after the
on-line filter code had been modified.}
\label{fig:6c1}
\end{figure}

For each event that passed all these cuts, the difference between the time of the 
positron scintillator cluster hit and the high-energy photon shower, corrected for the
different propagation distances of the positron and the photon, was calculated.  
Time differences were saved in 12
histograms -- keyed by photon spectrometer layer 1, 2, or 3, positron spectrometer upstream
or downstream, and
positron scintillator even or odd read out -- for contiguous groups of data runs that were
taken under similar
conditions.  Figure \ref{fig:6c1} shows time difference spectra for
coincidences between the
downstream, even positron scintillator TDCs and photon spectrometer layer 2 for two
different groups of
$\mu \to e\gamma$ data runs that were taken during
1994.  The two groups
have very different statistics because code was installed in the on-line filter
to enrich the data sample of ring counter,
high-energy photon coincidence
events during the middle of the 1994 data acquisition.
This code saved all high-energy photon events to tape 
that appeared to originate in the vicinity of a struck ring counter 
when propagated back to the $z$ axis.  This was independent of whether or not
the events also passed the on-line $\mu \to e\gamma$ event filter requirements.

The 12 histograms for each group of runs determined the 12 global timing
offsets needed to align
measurements in the positron and photon spectrometers.  All
of the final offset corrections were found to be less than 1.2 ns, which implies
that the timing calibration constants used during the on-line and preliminary off-line
filtering were sufficient to ensure good efficiency for acceptance of any true $\mu \to
e\gamma$ events.  The measured offsets were also quite stable over the duration of
each run period.  In principle, the 12 global timing offsets could
be reduced to 6 linearly independent time differences.  Attempts to do so,
however, found non-statistical effects at the level of $\sim$\,0.1 ns that were
attributed to small electronic propagation time differences through (nominally
identical) parallel circuit paths within the routing box and data acquisition gating
circuitry.  Therefore, the 12 separate offsets were used without modification during
the final stages of off-line analysis.

\subsection{$\mu \to e \gamma \nu \bar{\nu}$ Studies}
\label{sect:ib}

\begin{figure}[tbp]
   \begin{center}
     \mbox{\epsfxsize=3.3in \epsfbox{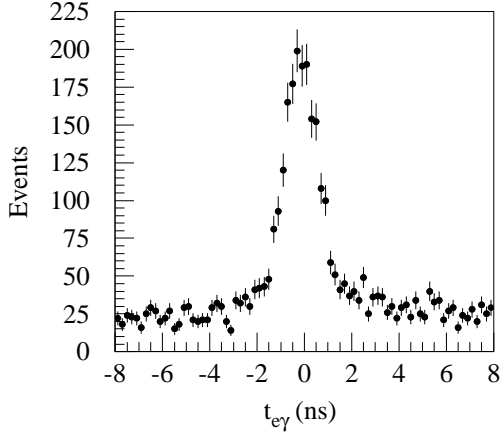}}
   \end{center}
   \caption{Values for $t_{e \gamma}$ from the process
   $\mu^{+} \rightarrow e^{+} \gamma \nu \overline{\nu}$ under the conditions
   of reduced rate and magnetic field.}
   \label{fig:6d1}
\end{figure}

      Observation of the IB process $\mu \to e \gamma \nu
\bar{\nu}$ demonstrates that the apparatus could detect
coincident $e- \gamma$ events.  At nominal beam intensity, this process was
completely engulfed by random coincidences.  Figure \ref{fig:6d1}
shows the $t_{e \gamma}$
spectrum for IB events using the standard first- and second-stage
hardware triggers, but with the beam intensity reduced by a factor of 60,
the magnetic field lowered by 25\%, and the $\mu \rightarrow e \gamma$
on-line filter suppressed.  The peak shown is for all energies of the detected
decay products.  The area of the peak is very sensitive to the exact 
acceptances
of the detector at its thresholds, and was calculated by MC simulation
to better than a factor of two.  If the data and the simulation are restricted to
$E_{\gamma} >$ 46 MeV, $E_{e} >$ 40 MeV,
and $\theta_{e \gamma} >120^{\circ}$,
the branching ratio is reproduced within 20\%.  The uncertainties in the IB
normalization do
not affect the precision of the $\mu \rightarrow e \gamma$ acceptance,
however, because the IB preferentially occurs near the energy-cut boundaries while the
$\mu \rightarrow e \gamma$ process occurs well above these cuts.

      The shape of the timing peak is characterized by a Gaussian with
0.77 ns rms.
The dominant contributor to the width is the photon timing, as measured in a stopping-pion
experiment, which must be scaled down from about 70 to 40 MeV for 
comparison.  At 52.8 MeV, the MC simulation indicated that the
photon-positron resolution was 0.68 ns rms.

\begin{figure}
 \begin{center}
     \mbox{\epsfxsize=3.3in \epsfbox{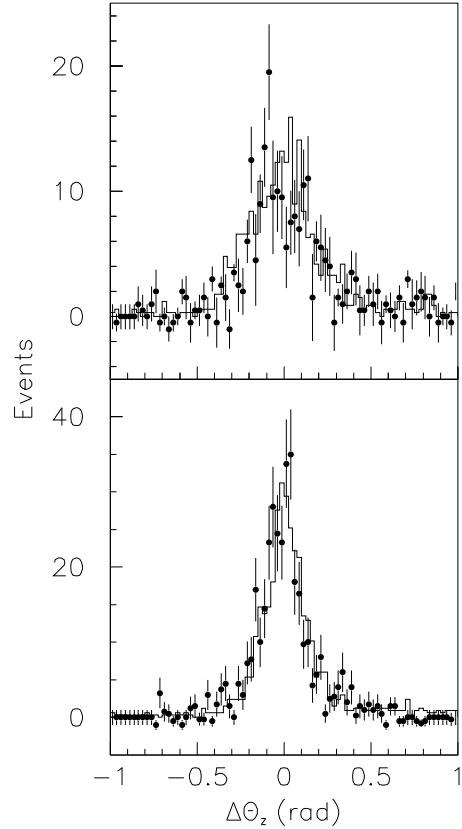}}
 \end{center}
\caption{The upper (lower) panel is the photon traceback resolution
$\Delta \theta_z$ for the 
inner (outer) conversions obtained from IB events where both the muon 
decay and photon conversion points were known.  Accidental 
coincidences have been subtracted.}
\label{fig:6d2}
\end{figure}

          In the IB and $\mu \rightarrow e \gamma$ processes, the 
origin of the photon is defined to be the intersection of the positron track with the 
target. The photon trace-back angle, $\Delta \theta_{z}$, specifys the difference
between the polar angles of the photon as determined from the lines connecting
the decay point to the photon conversion point and the direction of the reconstructed
$e^{+}e^{-}$ pair. The resolution of $\Delta \theta_{z}$ is dominated by
multiple scattering of the pair in the lead converters. The observed response for
inner and outer conversion layers of the IB process is in excellent agreement
with the MC simulation, as seen in Fig.\@ \ref{fig:6d2}.
The trace-back resolutions appropriate for the $\mu \rightarrow e \gamma$
analysis are 0.067 and 0.116 rad rms for conversions in the outer and the
inner lead layers, respectively.

\subsection{$\pi^0$ Decay Studies}
\label{sect:pi0}

The energy resolution of the photon spectrometer was measured by observing
the reaction $\pi^-p \rightarrow \pi^0 n$, $\pi^0 \rightarrow \gamma\gamma$.
When the  $\pi^-$ is captured at rest and the two photons from the $\pi^0$ decay are 
back-to-back, the energies of
the photons are 54.92 and 82.96 MeV.  Since the 54.92 MeV photon is
very close in energy to the 52.8 MeV photon from $\mu \rightarrow e \gamma$
decay, it was used to determine the energy resolution of the spectrometer
at 52.8 MeV.  

The $\pi^-$ were stopped in a CH$_2$ target and events with a potential
$\pi^0 \rightarrow \gamma \gamma$ pair in the spectrometer were collected
by triggering on events that had approximately back-to-back
high-energy photon first-level triggers in two different
spectrometer layers.  These events were
then analyzed by a filter that did a quick reconstruction of events and kept only
those that had at least one photon that reconstructed with an energy between 76
and 96 MeV.  These events were then analyzed by the full
reconstruction code and the opening angle of the two photons and the reconstructed
energies recorded.  The expected energies of the two photons were also
calculated from the opening angle of the photons and those energies 
recorded.  The energy resolution was determined by plotting the
difference between the measured energy and the calculated energy based
on the opening angle, after an opening angle cut was placed on the events.
For inner conversion layer events, the cut was $\theta_{{\gamma}{\gamma}} > 171^{\circ}$ and
for outer conversion layer events, it was $\theta_{{\gamma}{\gamma}} > 173.5^{\circ}$.
The cuts were selected to minimize resolution degradation due to uncertainties
in the opening angle, but still retain enough events to determine
the response functions accurately.  For each set of events, the error in the
calculated energy that came from the finite opening angle resolution was small
compared to the resolution of the measured photon energy.

\begin{figure}[tbp]
   \begin{center}
     \mbox{\epsfxsize=3.3in \epsfbox{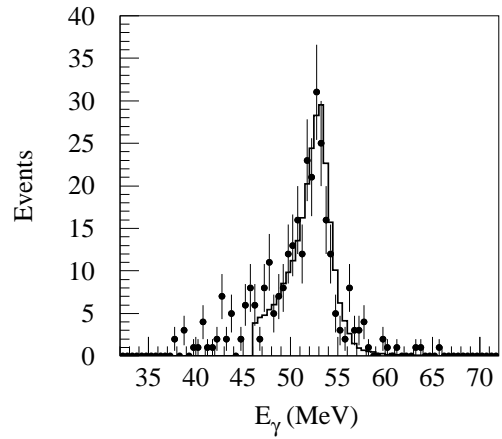}}
   \end{center}
  \caption{The energy resolution of reconstructed photons from pion decay
   for events that converted in the inner conversion layer.  The points
   come from data and the solid line shows reconstructed MC
   events. \label{fig:6e1}}
\end{figure}

\begin{figure}[tbp]
   \begin{center}
     \mbox{\epsfxsize=3.3in \epsfbox{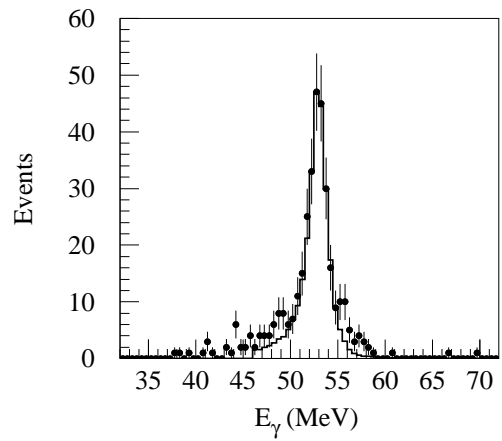}}
   \end{center}
  \caption{The energy resolution of reconstructed photons from pion decay
   for events that converted in the outer  conversion layer.  The points
   come from data and the solid line shows reconstructed MC
   events. \label{fig:6e2}}
\end{figure}

The energy resolutions that were obtained, for inner and outer conversion
layer events, are shown in Figs.\@ \ref{fig:6e1} and \ref{fig:6e2}, respectively,
where the measured energies have been shifted down by 2.1 MeV so
the peaks may be compared to the simulated $\mu \to e\gamma$
decay signal.
As can be seen, the energy resolution
for inner conversion layer events is significantly worse than the resolution
for outer conversion events.  Degradation occurs because the $e^+e^-$ pairs from
inner conversion layer events suffered significant multiple scattering
and energy loss in the outer lead conversion layer before they were
tracked by the drift chambers.  The central energy and width of the distributions were well
reproduced by the MC.  The differences in the low-energy tails were due to charge exchange of
in-flight pions from carbon in the CH$_2$ target.  The additional high-energy tails in the data
were associated with contributions from other opening angles, due to special difficulties
identifying the conversion point for the 83.0 MeV photon.  
The energy resolutions were 5.7\% and 3.3\% (FWHM) at 52.8
MeV for conversions in the inner and outer lead layers,
respectively.

The timing resolution of the photon spectrometer was determined by looking at the time
difference between the two photons in pion events.
The resolution from the time difference of the
two photons was divided by $\sqrt{2}$ and resulted in a timing resolution of 0.57 ns rms
for single photons.  

Finally, reconstruction inefficiencies caused by electronic problems that were not modeled in
the MC were studied with the pion data.  Data events were selected that included a reconstructed
high-energy (~82.96 MeV) photon, and had an angular difference of  $> 150^{\circ}$ between
the high energy photon and the center of the shower of the second photon that triggered the
event.  MC pion events were generated and passed through the pion trigger software with
the same analysis cuts applied.  The number of events that included a second reconstructed
photon in the
data set was compared to the number of photons reconstructed in MC yielding a correction
factor of 0.84 that was applied to calculated MC reconstruction efficiencies.

%\section{$\mu \to \lowercase{e} \gamma$ Data}
\section{Data Analysis}
\label{sect:mutoeg}

\subsection{Event Selection}
\label{sect:offline-filt}

The data recorded on line, $\:4.5\:\times\:10^{8}$ stored events, 
were passed through a set of computer
programs that reconstructed all events that the pattern recognition algorithms could
interpret. The analysis began by matching a run to constants that had been collected in
run-number-keyed data-base files.  These files included pedestals and timing
calibration constants created during the data acquisition.  The constants were used to
convert raw information from the electronics into physical quantities.  A separate set
of files contained the geometry of the apparatus.  Analysis of a subset of the data was
used to produce a list of dead wires in the positron chambers. 

The next step was to carry out the photon reconstruction, because this program was
much faster than the positron reconstruction.  Roughly 8\% of the original sample passed
energy and quality cuts in this
part of the code. Next the positrons were analyzed. About 30\% of the events that
passed the photon cuts could be reconstructed.  Very loose kinematic cuts were
placed on these events.  This left 0.15\% ($\:6.7\:\times\:10^{5}$) of the original
events that were output to data summary tapes.  

The above process took roughly one year of computing on a farm (20 processors) of
UNIX workstations.  These workstations were controlled by cshell scripts whose
purpose was to find a run of data that had not been analyzed and make the run
available to the processor.  The scripts reduced the processing to keeping the disks
filled with unprocessed data and checking the quality of the results.  Pass rates and
histograms corresponding to 40 different data distributions were monitored once per
tape.

In the final analysis of the remaining events, the tracking code stepped the positrons through the
magnetic field to get the
best position and momentum vector at the target.  
Improved timing constants described in
Sect.\@ \ref{sect:offline-timing} were also
incorporated at this time.  The results were added to the data stream and new data
summary tapes were produced.  Additionally, PAW ntuples \cite{PAW} for 60
variables were retained.  These ntuples were useful for obtaining high statistics
distributions of random, uncorrelated backgrounds that were needed for the likelihood
analysis.

With the final kinematics calculated for each event, the data set was further
reduced to $\:5.5\:\times\:10^{3}$ events that were fully reconstructed and
of continued interest.  These events were required to have good-quality positron
reconstructions, to satisfy separate $\chi_{\nu}^2$ cuts on the positron and photon
fits, and to pass loose cuts on the signal kinematics ($E_{e}\:>\:$50 MeV,
$E_{\gamma}\:>\:$46 MeV, $|t_{e \gamma}|\:<\:$4 ns, 
cos($\theta_{e \gamma}$)$\:<\:$-0.9962, and $|\Delta \theta_{z}|\:<\:$0.5 rad).
Events in which the positron momentum vector at the decay point appeared to lie
within 5$^{\circ}$ of the plane of the target were discarded.

To remove incorrectly reconstructed events, the images of the photon showers in the
pair spectrometers were manually scanned.  The efficiency for real photons
was monitored by mixing about 500 MC 52.8 MeV photon events into the
sample in a non-identifiable way and finding that 91\% of the MC events
passed, whereas only 73\% of the data events were selected.  Most of the rejected data
consisted of two overlapping low-energy photon showers
that had been reconstructed by the analysis program as a single high-energy shower. 
The remaining sample of 3971 data and 450 MC events
was stored in an ntuple as the input to
the likelihood analysis (Sect.\@ \ref{sect:likelihood}).
This sample was large enough to allow a study of the
background.

\subsection{Normalization}
\label{sect:normalization}

Two quantities that are very important for the calculation of the 
$\mu \rightarrow e \gamma$ branching ratio are the acceptance of the detector
and the number of muons stopped in the target.  The number of stopped muons 
was determined by counting the number of decay positrons striking one of the
scintillators in the upstream array of the positron scintillator detectors, as described in
Sect.\@ \ref{sect:beam}.  
The acceptance of the apparatus, which includes geometrical, trigger, and
pattern recognition constraints, was obtained by simulating
$\:1.2\:\times\:10^{7}$ unpolarized $\mu^{+} \rightarrow e^{+} \gamma$ decays
for a 1993 data set that was then used as the standard run.
Of the thrown MC events, $5.2\:\times\:10^{4}$ survived processing by the same
codes used for the data analysis.  Thus the probability for detection of a $\mu
\rightarrow e \gamma$ decay was $\:4.3\:\times\:10^{-3}$. 
This value included a 9\% reduction for the inefficiency of manual scanning.  
An additional 20\% reduction was made to account for inadequacies in the MC
simulation that over-estimated the acceptance.  The MC shortcomings 
primarily involved inter-channel cross talk and were estimated to contribute only 4\% to the
overall uncertainty in the acceptance by comparing the images of many data and MC events.  

Since the data were taken over a span of 3 years, the efficiencies of
the detector underwent a number of changes.  To account for this, run-by-run 
corrections were made to the stopped muon rate relative to the standard run.  
The nearly 4000 data runs
were divided into several groups that reflected major changes in the detector performance
such as an inoperative photon layer, an inoperative positron chamber, a
change in the instantaneous beam rate, or a change in the noise level of the photon
spectrometers.  The largest change, up to 40\%, occurred when a
photon layer was inoperative.  For the '94 and '95 data there was an additional
10-15\% loss in performance due to a reduced acceptance in the first drift chamber, DC1. 
This reduction occurred because the drift chamber threshold was lowered and led to increased
inter-channel cross talk, which contaminated the vertex and edge of some events.
The rate dependence of the positron reconstruction algorithm also led to a change in
efficiency when the instantaneous beam rate changed.  The majority of the data were
taken at a beam rate within $\pm$\,12\% of the nominal 250 MHz.  Fluctuations in the
instantaneous beam rate resulted in a correction of approximately 2-3\%.  The 
variation in the muon stopping rate had less than a 1\% effect on the photon
reconstruction efficiency and no corrections were made for it.  The
high instantaneous beam rate was also responsible for increased dead time in the 
detector and resulted in a loss of about 10\% of the effective number of muons
stopped in the target.

In addition to the corrections noted above, there were other run-by-run
corrections of order 5\% or less for changes such as modifications to the software
used in data acquisition and analysis.  After all of the above corrections were made,
the estimated total number of muons stopped
in the target during the 3 calendar years
of data taking (8 $\times$ 10$^6$ s of live time) was $N_s$ = 1.2 $\times$ 10$^{14}$.
After convoluting  $N_s$ with the acceptance,
the single event sensitivity for the experiment was $\:2.3 \pm
0.2\:\times\:10^{-12}\:$=$\:1/N_{\mu}$, where $N_{\mu}$ was the number of useful
stopped muons.

\subsection{Cut Analysis}
\label{sect:box}

\begin{figure}
 \begin{center}
     \mbox{\epsfxsize=3.3in \epsfbox{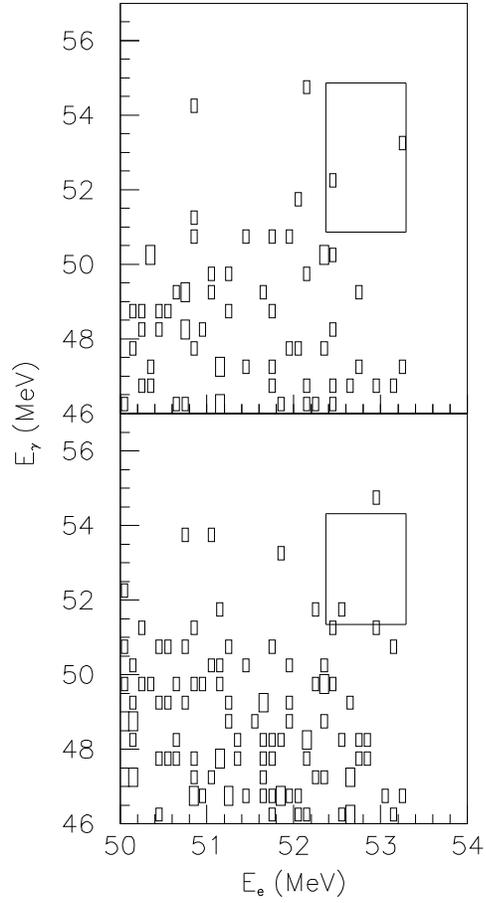}}
 \end{center}
\caption{A box analysis of the data.  The upper (lower) panel 
shows the events from 
inner (outer) conversions plotted as a function of the photon and positron 
energies.  The large rectangle in both panels is the 
$\mu \to e\gamma$ signal area for $\pm$\,2\,$\sigma$ in the parameters shown.}
\label{fig:8_1}
\end{figure}

Before embarking on 
the complexities of a likelihood analysis, the result of a so-called
blind box analysis is presented.  In this method, cuts are placed around the signal 
region based on the width of the response functions.  The width of the 
window was taken to be $\pm$\,2$\sigma$ in each of the five parameters, 
$E_{\gamma}$, $E_{e}$, $t_{e \gamma}$, $\theta_{e\gamma}$ and 
$\Delta \theta_{z}$.  As there 
were many subsets of the data for different types of events, it is imprecise 
to combine all the data together.  Nevertheless, the results are plotted in 
Fig.\@ \ref{fig:8_1} with the
$E_{\gamma}$ and $E_e$ variables shown explicitly for inner and 
outer conversions.  The box in each panel is where the signal would be 
found in the absence of background.  Three events fall inside the 
boxes, but they appear consistent with the background.  If one takes 
the three events inside the boxes as background, then the branching ratio limit is 
about 20\% greater than the one obtained by the likelihood analysis below.  
The primary difference arises because each of the events is near the edge 
of the boxes and is improbable as real signal.

\subsection{Likelihood Analysis}
\label{sect:likelihood}

The five kinematic properties, in conjunction with the detector response, 
determined the likelihood that a $\mu^{+} \rightarrow e^{+} \gamma$ 
signal was detected. The determination of the 
detector acceptance and response functions relied on MC simulation to 
extrapolate from experimental input 
for measured responses to the kinematic region of the $\mu \rightarrow e \gamma$ signal.  The
determination of the number of $\mu \rightarrow e \gamma$
events in the sample was evaluated using the likelihood method described in the
analysis of previous experiments~\cite{kinnison82}. The formula for the
normalized likelihood is
\begin{eqnarray}
{\cal L}(N_{e \gamma},N_{IB})  = 
\left( 1-\frac{N_{e \gamma}}{N}-\frac{N_{IB}}{N} \right)^{N-m} \nonumber \\
\times \prod_{i=1}^{m} \left( \frac{N_{e \gamma}}{N} \left( \frac{P}{R}-1
\right)
+\frac{N_{IB}}{N} \left( \frac{Q}{R}-1 \right)
+1 \right),
\label{eqn:like1}
\end{eqnarray}
where $N=3971$, $N_{e \gamma}$ is the number of signal events, $N_{IB}$ is the
number of IB events, and $P$, $Q$, and $R$ are the probability density functions
(PDF) for signal, IB, and randoms of each of the five parameters describing the event. 
The value of $m \le N$ was chosen to speed the calculation and leave the value of the
likelihood unchanged if $P/R \ll 0.01$ and $Q/R \ll 0.01$ for all events left out of the
product.  The value of $m$ was 3832 for this experiment.  The events fell into the
following categories:  positron topology, photon conversion plane, target intersection
angle, and data collection period. As a result, PDFs were extracted for each class of
events and applied according to the classification of individual events.

\begin{figure}
 \begin{center}
     \mbox{\epsfxsize=3.3in \epsfbox{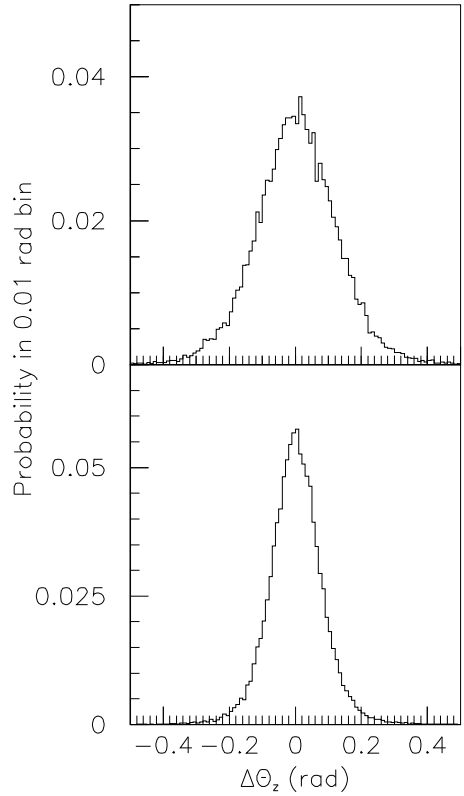}}
 \end{center}
\caption{The signal PDF for $\Delta \theta_z$ for inner 
(outer) conversions is shown in the upper (lower) panel, as generated by MC 
simulation.}
\label{fig:8_2}
\end{figure}

\begin{figure}
 \begin{center}
     \mbox{\epsfxsize=3.3in \epsfbox{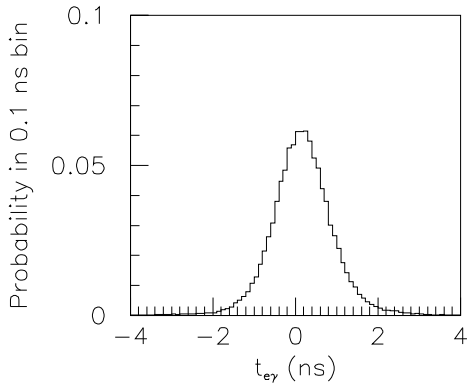}}
 \end{center}
\caption{The PDF for the timing variable associated with the 
$\mu \to e\gamma$ signal.  This PDF was generated with MC that reproduced the 
experimental IB relative-time distribution, and had a slight dependence on 
the photon energy.}
\label{fig:8_3}
\end{figure}

\begin{figure}
 \begin{center}
     \mbox{\epsfxsize=3.3in \epsfbox{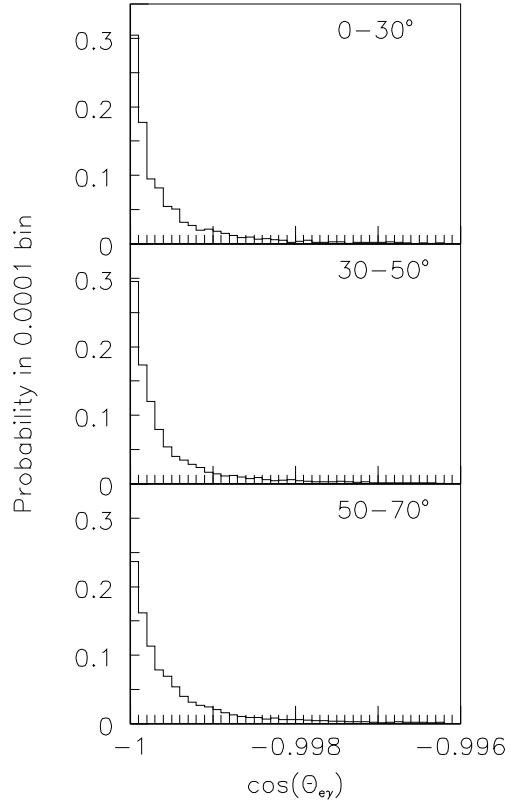}}
 \end{center}
\caption{The PDF 
function for the cosine of the angle between the positron and the gamma ray, 
as generated by MC simulation for the $\mu \to e\gamma$ signal.  The panels are 
labeled by the angle between the target normal and the positron momentum 
unit-vector at the target, which can range from 0-180$^{\circ}$.  The PDFs are 
slightly asymmetric about 90$^{\circ}$ because the target faced a dwarf on one side 
and faced the gap between dwarfs on the other side, which induced a small 
change in the precision of measuring the positron trajectory.}
\label{fig:8_4}
\end{figure}

The PDFs $P$ and $R$ are the products of statistically independent PDFs for the five
parameters, each normalized to unit probability over the full range of the variable for
the sample.  The signal distributions for the photon depended on whether the photon
converted in the inner or outer lead convertor in each pair spectrometer.  The
distributions for $E_{\gamma}$ shown in Figs.\@ \ref{fig:6e1}
and \ref{fig:6e2} were derived from
MC simulations that had the same input as those compared to the pion
data in Sect.\@ \ref{sect:pi0}.  The energy resolutions were 3.3\% and 5.7\%
(FWHM) for conversions in the outer and inner lead layers, respectively.  Likewise, the
distributions for $\Delta \theta_{z}$ were derived from a MC simulation that had the
same input as those compared to the IB data in Sect.\@ \ref{sect:ib}.  They are plotted
in Fig.\@ \ref{fig:8_2}.  The trace-back angular resolutions were 
0.067 and 0.116 rad rms.
These distributions were somewhat narrower than those measured from 
the IB in Fig.\@ \ref{fig:6d2} because the photon energy is higher in the
$\mu \to e\gamma$ 
process than in the IB.  The distribution for $t_{e \gamma}$ was 
also based on the IB data and had a resolution of 0.68 ns rms as shown in
Fig.\@ \ref{fig:8_3}.  There was no way to measure the response function for
cos($\theta_{e \gamma}$).  Thus the MC simulation was relied upon to produce this
distribution and gave the average FWHM for cos($\theta_{e \gamma}$) as
$1.21\:\times\:10^{-3}$ at $180^{\circ}$.  Given helical tracks, the location
of the target was critical to obtaining the correct absolute value of 
cos($\theta_{e \gamma}$).  The mechanical survey provided the most accurate
measurement for the analysis.  The resolution also depended
on the angle the track made with the target, and some representative distributions
are plotted as a function of target angle in Fig.\@ \ref{fig:8_4}.  All of the above
distributions were used directly in the likelihood analysis and came from
high-statistics MC simulations.

\begin{figure}
 \begin{center}
     \mbox{\epsfxsize=3.3in \epsfbox{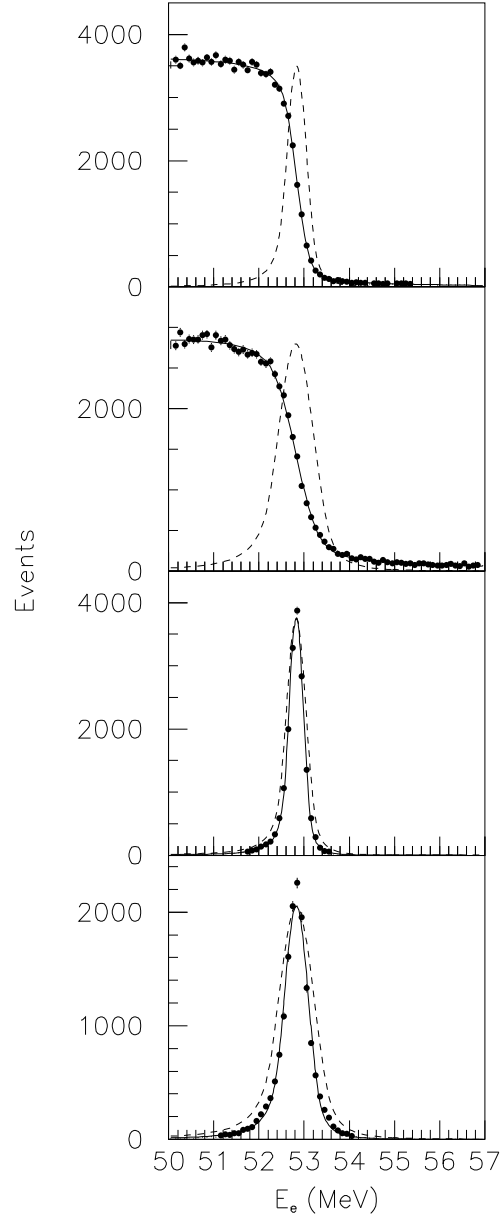}}
 \end{center}
\caption{Unnormalized PDFs for the positron energy 
distributions in correspondence to the cases described in Table V.  The top 
two panels illustrate the data for the last part of 1995 for groups A and 
C.  The solid lines are fits that are normalized to become the background 
PDFs.  The dashed lines are the monoenergetic line shapes produced with the 
same parameters as the fit, and they are normalized to become the signal 
PDFs.  The data points in the lower two panels are generated with 
MC simulations.  The solid curves are fits to these line shapes 
used to extract the centroids and resolutions for the table.  The dashed 
lines are the line shapes produced from the fits in the upper two panels; 
they illustrate the energy resolution degradation described in the text.}
\label{fig:8_5}
\end{figure}

Three types of positron topologies (i.e., number of dwarfs encountered and
number of loops) were determined empirically from the 1993 data to obtain three
statistically equivalent size samples.  When extended to all the data, the 
groups maintained their resolution characteristics.  For the three types of positron
topologies, the signal was extracted by fitting the positron edge in the following way.  
The function,
\begin{equation}
$${\cal F}(E) = \cases{\frac{C_L}{(E_0-E)^{2.05}}&if $E < E_0-
\sqrt{2.05} \sigma$, \cr
ae^{-\left( \frac{E-E_0}{\sqrt{2} \sigma} \right)^2}&if $E_0-
\sqrt{2.05} \sigma \le E \le E_0+ \sqrt{3} \sigma$, \cr
\frac{C_U}{(E-E_0)^3}&if $E_0+ \sqrt{3} \sigma < E$, \cr}$$
\label{eqn:like2}
\end{equation}
represented the positron monoenergetic line shape.
The constants $C_L$, $a$, and $C_U$ were related by the continuity condition 
at the region boundaries.  The asymmetry in the non-Gaussian tails reflected the 
straggling present in the energy loss of the positrons.
The function ${\cal F}(E)$ was convoluted with the function,
\begin{equation}
$${\cal G}(E_e-E_0) = \cases{1+b(E_e-E_0)&if $E_e \le E_0$, \cr
c+d(E_e-E_0)&if $E_0 < E_e$, \cr}$$
\label{eqn:like3}
\end{equation}
and then fit to the spectrum.  The terms involving $b$ through $d$ were needed to
accommodate the change in acceptance as a function of energy below $E_0$, and
ghost events above $E_0$.  The parameters of the fit were $a$ through $d$, $E_0$,
and $\sigma$.  The values of the fit parameters were used in the likelihood
analysis for events by topology group and run period.  The results were quite stable
as a function of run period.  Some representative results for $E_0$ and $\sigma$
for 1995 are shown in Table \ref{tab:81}.
\begin{table}[tbp]
\caption{Representative fitting parameters for topology groups $A$ and $C$ for
the last third of the 1995 data, one of the periods treated separately in the
likelihood analysis.
\label{tab:81}}
\begin{tabular}{lcc}
 &$E_0$ (MeV)&$\sigma$ (MeV) \\ \tableline
Group $A$&52.835(0.019)&0.218(0.009) \\
Group $C$&52.814(0.030)&0.391(0.018) \\
Group $A$ MC&52.834(0.002)&0.161(0.001) \\
Group $C$ MC&52.826(0.003)&0.270(0.003) \\
\end{tabular}
\end{table}
All the central energies agreed within errors to $m_{\mu}/2$.  The pattern of
resolutions between MC and data agreed, but the data resolution was clearly degraded by
problems with electronic noise in the detector that smeared the location of the space
points.  This was not simulated in the MC.  The experimental values for $\sigma$
were used for the signal distributions in the likelihood analysis.  The fits are displayed
in Fig.\@ \ref{fig:8_5}, where the solid curves are the fits to the points and the
dashed line is the parametrized line shape used for the $\mu \rightarrow e \gamma$ signal PDF. 
Figure \ref{fig:5b2} is a similar plot for group $B$ data for
all years.

\begin{figure}
 \begin{center}
     \mbox{\epsfxsize=3.3in \epsfbox{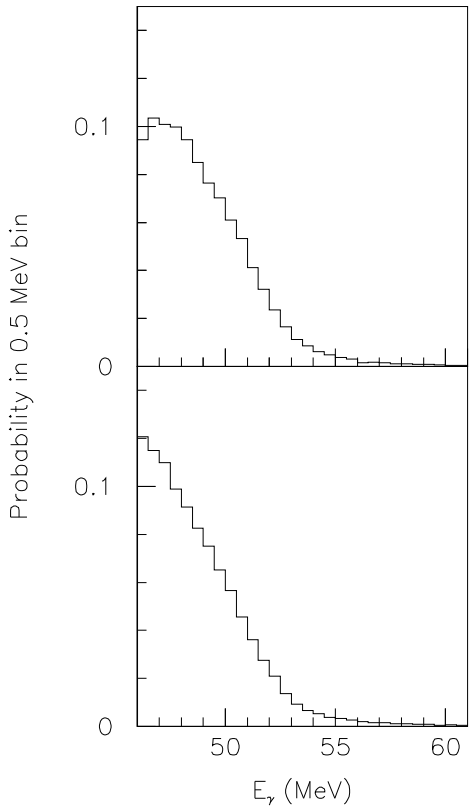}}
 \end{center}
\caption{The 
upper (lower) panel shows the energy PDF of the background gamma rays for the 
inner (outer) gamma ray conversions.}
\label{fig:8_6}
\end{figure}

\begin{figure}
 \begin{center}
     \mbox{\epsfxsize=3.3in \epsfbox{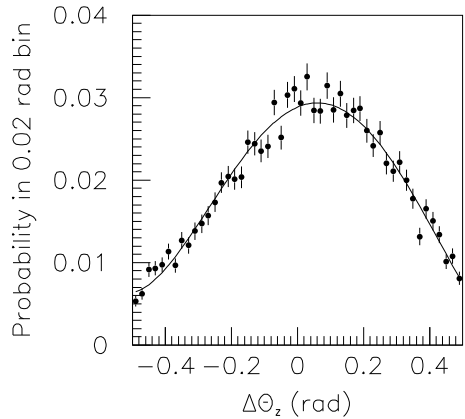}}
 \end{center}
\caption{A representative background PDF for 
$\Delta \theta_z$ generated by examining events out of time coincidence.}
\label{fig:8_7}
\end{figure}

\begin{figure}
 \begin{center}
     \mbox{\epsfxsize=3.3in \epsfbox{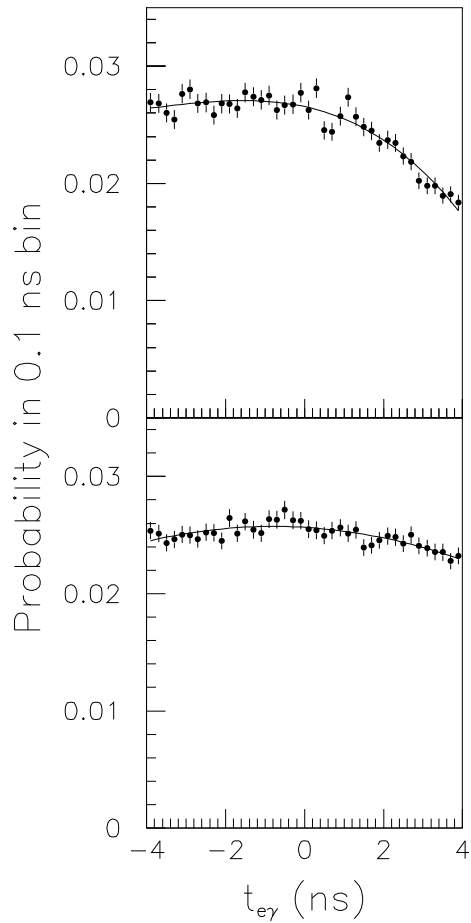}}
 \end{center}
\caption{The 
accidental-background PDF for the relative time between the
positron and photon for 1993 (1994-95) is shown in the upper (lower) 
panel.  The smooth curve through the data was used in the likelihood 
analysis.}
\label{fig:8_8}
\end{figure}

The background PDFs were extracted from the spectral shapes of a much larger
sample of events, where the other statistically independent
parameters remained loosely constrained.  The PDFs for the outer and inner photon
conversions are shown in Fig.\@ \ref{fig:8_6}, where the shape was modified to
account for the efficiency of the hand scanning.  The correction functions used
were:
\begin{equation}
$$\cases{0.77&if $E_{\gamma} \le 51$ MeV, \cr
0.77-0.064(E_{\gamma}-51)&if $51$ MeV $< E_{\gamma} \le 56$ MeV, \cr
0.45&if $56$ MeV $< E_{\gamma}$, \cr}$$
\label{eqn:like4}
\end{equation}
and:
\begin{equation}
$$\cases{0.71&if $E_{\gamma} \le 51$ MeV, \cr
0.71-0.086(E_{\gamma}-51)&if $51$ MeV $< E_{\gamma} \le 56$ MeV, \cr
0.28&if $56$ MeV $< E_{\gamma}$, \cr}$$
\label{eqn:like5}
\end{equation}
for outer and inner photon conversions, respectively.
The error in Eqs.\@ \ref{eqn:like4} and \ref{eqn:like5} was limited by
the statistics of the sample of events that were manually scanned.  Empirically, the
likelihood function seemed moderately insensitive to the coefficients in these
equations.
A representative PDF 
for $\Delta \theta_z$ is shown in Fig.\@ \ref{fig:8_7}.  The distribution was not very 
dependent on whether the conversion was from the inner or outer  
lead converters, but did have a mild dependence on the photon energy.
This shape was largely determined by the beam stopping
distribution.  Although this 
distribution is peaked at $\Delta \theta_z = 0$, it is still significantly 
broader than those in Fig.\@ \ref{fig:8_2} and gave some useful discrimination between 
prompt and accidental processes.  The PDFs for $t_{e \gamma}$ are shown in Fig.\@
\ref{fig:8_8}.
They are not quite flat because of the time-dependent properties of the on-line filter.
For the data taken in 1993, the filter was set for three timing loops in the
ARC algorithm (see Sect.\@ \ref{sect:arc}) and caused a loss of efficiency beginning near
$+2$ ns.  This effect was noticeably reduced for the
balance of the data.  The
background PDF for cos($\theta_{e \gamma}$) was a constant for all target
intersection angles.  The PDFs for $E_e$ were obtained by normalizing the fits
to the muon-decay spectrum, examples of which are shown in Figs.\@ \ref{fig:5b2}
and \ref{fig:8_5}.

In Eq. \ref{eqn:like1}, $Q$ was taken from MC simulation of the IB and had
correlations amongst the variables $E_{\gamma}$, $E_{e}$, and 
$\theta_{e \gamma}$.  The PDFs for $t_{e \gamma}$ and
$\Delta \theta_{z}$ were statistically independent and were obtained in an 
analogous way to $P$, except that the response was for a
lower mean energy photon.  The correlated matrix element varied by more than three
orders of magnitude over the range of the likelihood function.  If calculated by
randomly sampling events over the entire range of the likelihood function, there
would have been no statistics at the highest energies.  Hence, it was calculated by
convoluting the MC simulation of the IB process with the detector acceptance in small
regions of the parameter space and weighting the passing events by the integrated 
IB matrix element for that region.  The regions and branching ratios are given in
Table \ref{tab:82}.  The overall PDF was normalized to unity. 

\begin{table}[tbp]
\caption{The branching ratios for the bins of the simulation of the 
correlated IB matrix
element.  All kinematically allowed $\theta_{e \gamma}$ are encompassed.
The energies are in MeV.
\label{tab:82}}
\begin{tabular}{lllllllllll}
&&&&& \multicolumn{2}{c}{$E_e$} & \\
%&&&&&&$E_e$&&&& \\
&&&49-50&&50-51&&51-52&&52-53 \\
&53.0 \\
&\@ \@ $\vert$&&$2.84^{-11}$&&$1.22^{-11}$&&$3.58^{-12}$&&$3.36^{-13}$ \\
&50.5 \\
&\@ \@ $\vert$&&$1.76^{-10}$&&$8.02^{-11}$&&$2.49^{-11}$&&$2.44^{-12}$ \\
$E_{\gamma}$&48.0 \\
&\@ \@ $\vert$&&$4.74^{-10}$&&$2.21^{-10}$&&$7.07^{-11}$&&$7.01^{-12}$ \\
&45.5 \\
&\@ \@ $\vert$&&$9.65^{-10}$&&$4.53^{-10}$&&$1.46^{-10}$&&$1.46^{-11}$ \\
&43.0 \\
\end{tabular}
\end{table}

\begin{figure}
 \begin{center}
     \mbox{\epsfxsize=3.3in \epsfbox{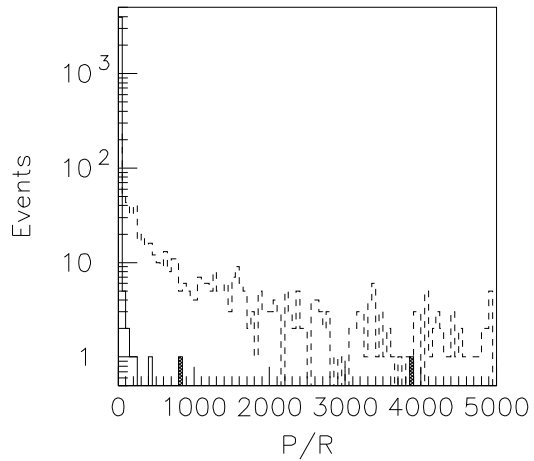}}
 \end{center}
\caption{The $P/R$ 
distributions for the data (solid line) from this experiment, and for the 
MC predicted signal (dashed line) for this experiment.  The two 
hatched events were added to the data sample
when the cuts were relaxed as discussed 
in the text.}
\label{fig:8_9}
\end{figure}

The likelihood function evaluated the statistical separation between signal,
IB, and background.  Figure \ref{fig:8_9} shows the value of $P/R$ for the data
sample and for a set of MC events.  The data largely had values below one but had a
tail that extended out to around 300.  The signal was peaked near one but had most of
its area at values greater than 300.  The values ran up to near 12,000 but are plotted
only to 5,000.  The interpretation of the overlap of the two plots is that the
measurement is not free of background.

To observe the impact of quality constraints in the pattern recognition, they were
relaxed to produce a sample three times larger. One event emerged with a large value
of $P/R$ that was significantly separated from the distribution. The value of $P/R$
for this event was 3888, which made it resemble signal.  Its kinematic properties were
$E_{\gamma}$ = 52.58 MeV, $E_e$= 53.37 MeV, $t_{e \gamma}$ = 0.028 ns,
$\Delta \theta_{z}$ = 0.068 rad, and cos($\theta_{e \gamma}$) = -0.99988.  In
addition, it was within 19$^{\circ}$ of being perpendicular to the target and had 11 of
19 possible triples plus the outgoing Snow White anode as part of the positron track. 
The positron energy was located in the region of the spectrum that was particularly
sensitive to the signal because the PDF for signal fell more slowly than the one for
background.  However, this event had a large positron $\chi_{\nu}^2$ =
4.90/degree-of-freedom, indicative of a ghost track.  Therefore, the event was
considered either unphysical or statistically insignificant, though if it were real, it
would correspond to a branching ratio at the single event sensitivity of the
experiment, $(2.3\,\pm\,0.2)\:\times\:10^{-12}\:$.  The adopted constraints produced a
sample with considerably less background. The result presented below was stable
against changes in the constraints; e.g., the higher value of $N_{e \gamma}$ was
compensated by a corresponding increase in acceptance.

\begin{figure}
 \begin{center}
     \mbox{\epsfxsize=3.3in \epsfbox{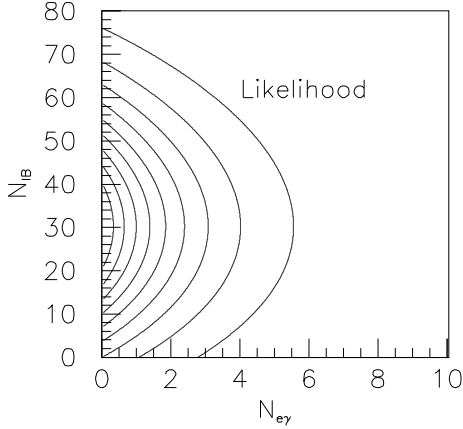}}
 \end{center}
\caption{A contour plot of the
two-dimensional likelihood 
function as a function of the numbers of prompt events of IB and
$\mu \to e\gamma$.}
\label{fig:8_10}
\end{figure}

A plot of the two dimensional likelihood function is shown in Fig.\@
\ref{fig:8_10}.
The contours of constant likelihood have their extrema along the
$N_{e \gamma}$ and $N_{IB}$ axes, demonstrating the statistical 
independence of the separation between these two processes.  The peak of the
likelihood function was at $N_{e \gamma}$=0 and $N_{IB}$=30$\pm$8$\pm$15.
The systematic error assigned to $N_{IB}$ was due to the uncertainty in the shape of
the background time spectrum when the events were filtered by the on-line program.
The expected number of IB events was 36$\pm$3$\pm$10, where the systematic
error was due to finite resolution effects across the cut boundaries. The 90\%
confidence limit is the value for $N_{e \gamma}$ where 90\% of the area of the
likelihood curve lies below $N_{e \gamma}$ and $N_{IB}$ is maximal.  This
value was $N_{e \gamma}<5.1$.  Therefore, the limit on the branching ratio is
\begin{equation}
\left. \frac{\displaystyle \Gamma(\mu^+ \rightarrow e^+ \gamma)}
{\displaystyle \Gamma(\mu^+ \rightarrow e^+ \nu \overline{\nu})} \right. \leq
\frac{\displaystyle 5.1}{\displaystyle N_{\mu}} = \:1.2\:\times\:10^{-11}\:
(90\%\:{\rm CL}).
\label{eqn:like6}
\end{equation}

\section{Goals vs.\@ Achievements}

The 90\%-confidence sensitivity for $\mu^{+} \rightarrow e^{+} \gamma$ for
this experiment analyzed with the likelihood method
is given by

\begin{equation}
{\cal S} =
\frac{N_{e\gamma}}{\left( \frac{\textstyle \Omega}{\textstyle 4 \pi} \right) \epsilon_e
\epsilon_{\gamma} 
N_{s}},
\label{eqn:goals1}
\end{equation}
where the terms in the denominator are the fractional solid angle, the 
detection
efficiency for each particle, and the total number of muon decays.  
The product of the first three terms in the denominator, neglecting
correlations, is roughly the acceptance of the apparatus.  These 
factors identify the major contributions to the acceptance and are given
in the second column of Table \ref{tab:91}.  The product has the value
$\:3.9\:\times\:10^{-3}$ and can be compared to the 
$\:0.8\:\times\:4.3\:\times\:10^{-3} = \:3.4\:\times\:10^{-3}$ given in 
Sect.\@ \ref{sect:normalization}.

The sensitivity at
which one event of background \cite{vanderschaaf97} should be seen is
\begin{equation}
{\cal B} =
\left( \frac{R_{\mu}}{d} \Delta t \right)
\left( \frac{\Delta E_e}{m_{\mu}/2} \right)
\left( \frac{\Delta E_{\gamma}}{15 m_{\mu}/2} \right) ^2
\left( \frac{\Delta \theta}{2} \right) ^2
f(\theta_{\gamma})\eta_{\rm IBV}.
\label{eqn:goals2}
\end{equation}
The terms are the muon stop rate divided by the beam duty factor multiplied
by the detector time resolution,
the positron energy resolution in reduced units, the photon energy resolution
in reduced units, the angular resolution, and the background reduction
factors from photon angle traceback and inner bremsstrahlung veto.  All
resolutions are FWHM.  The quadratic
dependence of the photon energy term is related to the bremsstrahlung 
nature of the energy spectrum, while that of the angular resolution
arises from phase space considerations.

     The proposal for the MEGA experiment \cite{proposal86} predicted a  
90\%-confidence
sensitivity for $\mu \rightarrow e \gamma$ as 
$9\:\times\:10^{-14}$.  Later engineering considerations reduced this 
to $4\:\times\:10^{-13}$, primarily because (1) the
solenoid could accommodate only three photon spectrometers instead of five,
(2) each photon spectrometer had only two rather than three lead converter
sheets (photon conversions in the innermost third sheet reconstructed
poorly), and (3) the achievable overall solid angle was 30\% smaller than
proposed.  Operation of the apparatus revealed a dramatically lower positron
reconstruction efficiency and a somewhat lower photon reconstruction efficiency
than expected.  These were attributed to electronic cross talk among the anode
and cathode readout channels of the positron spectrometer and among the photon
spectrometer delay line cathodes.

     The proposed sensitivity was essentially free of background.  However,
MC simulations performed after the detector was constructed
showed degradations in the resolutions to a background
free sensitivity of $1.6\:\times\:10^{-12}$.  These losses were largely 
due to inadequacies of the early-stage simulations, where measurements of the
responses of prototype detectors were not available.  In particular, the
scintillator time resolution was affected by the small number of
photoelectrons emitted from the cathodes of the photomultiplier tubes; the photon
energy resolution was affected by the use of delay lines rather than stereo
cathodes; and the positron momentum resolution was degraded by
changes in the geometry and wire configuration of the positron spectrometer.
%Furthermore, the absence of the proposed IBV scintillators
%on the upstream
%and downstream pole faces of the solenoid's iron yoke
%(see Fig.\@ \ref{fig:3a1})
%precluded rejection
%of background events containing an energetic inner bremsstrahlung photon
%(always accompanied by a detectable soft positron).
The single largest loss of background rejection capability came from the failure of the IBV
detectors to perform as expected. As noted in Sect.\@ \ref{sect:ibv}, the IBV scintillators that
were to line the upstream pole tip penetration were never installed.  However, the downstream
IBV detectors alone were expected to veto 40\% of the photons above 51 MeV originating from
$\mu \to e \gamma \nu \bar{\nu}$. But analysis of the IBV data demonstrated that only 3.5\% of
the high energy photons detected by the pair spectrometers were associated with IBV hits,
including 2\% real and 1.5\% random coincidences. This rate was essentially independent of
photon energy for $E_{\gamma} > 42$ MeV, rather than increasing with increasing
$E_{\gamma}$.  For these reasons, the inner bremsstrahlung veto was not used in the final
analysis.

     The changes in sensitivity and resolution between design and final
data analysis are given in Tables \ref{tab:91} and \ref{tab:92},
respectively.  In Table \ref{tab:92}, the degradation factor includes the
appropriate power from Eq.\@ \ref{eqn:goals2}.  The photon energy resolution
degradation factor is weighted by the fraction of events where the photon
converted in the inner or outer lead sheet.  In both tables, the total factor
is the product of the individual factors.

\begin{table}[tbp]
\caption{The contributions to the signal sensitivity of the MEGA experiment
at the design stage and after a complete analysis of the data.
\label{tab:91}}
\begin{tabular}{lccc}
\\
        &        &        &Degradation\\
Quantity&Designed&Achieved&Factor \\
\tableline
$N_{e\gamma}$ (90\% C.L.) & $\le 2.3$ & $\le 5.1$ & 2.2 \\
$\Omega/4 \pi$&0.42&0.31&1.4 \\
$\epsilon_e$&0.95&0.53&1.8 \\
$\epsilon_{\gamma}$&0.051&0.024&2.1 \\
$N_{s}$&$\:3.6\:\times\:10^{14}$&$\:1.2\:\times\:10^{14}$&3.0 \\ \tableline
Total Factor&&&34.9 \\
\end{tabular}
\end{table}

\begin{table}[tbp]
\caption{The contributions to the background sensitivity of the MEGA experiment
at the design stage and after a complete analysis of the data.
\label{tab:92}}
\begin{tabular}{lccc}
\\
        &        &        &Degradation \\
Quantity&Designed&Achieved&Factor \\ \tableline
$R_{\mu}$ (MHz)&30.0&15.0&0.5\\
$t_{e \gamma}$ (ns)&0.8&1.6&2.0 \\
$E_{e}$ (MeV)&0.25&0.54&1.5 \\
$E_{\gamma}$ (MeV)&1.7&1.7,3.0&1.6 \\
$\theta_{e\gamma}$ (deg)&1.0&1.9&3.6 \\
$\theta_{\gamma}$ (deg)&10.0&10.0&1.0 \\ 
$\eta_{\rm IBV}$ & 0.2 & 1.0 & 5.0 \\ \tableline
Total Factor&&&43.3 \\
\end{tabular}
\end{table}

     The degradation factors in Tables \ref{tab:91} and \ref{tab:92} arose
principally from three phenomena.  First, electronic cross talk in the positron
spectrometer limited the muon stop rate, reduced the reconstruction
efficiency, and degraded the energy and angular resolutions of the
reconstructed tracks.  Second, cross talk in the photon spectrometer delay line
cathodes reduced the photon reconstruction efficiency and degraded the
energy and conversion point resolutions.  Finally, an accident in 1993,
where the photon spectrometer was dropped from a crane at a height of about
30 cm, led to eventual crazing of the scintillators and a subsequent reduction
of light output; this loss of light reduced the scintillator efficiency and 
worsened the positron-photon timing resolution.

\section{Conclusion}

A high-precision search for the rare muon decay mode
$\mu^+ \to e^+ \gamma$ has been performed with the
MEGA detector.  A
maximum-likelihood analysis of the data established
a new upper limit for
the branching ratio of ${\cal B}(\mu^+ \to e^+ \gamma) <
1.2\,\times\,10^{-11}$ with 90\% confidence.  This upper limit
constrains the existence of physics
outside the standard $SU(3) \times SU(2) \times U(1)$
model of the strong and electroweak interactions, since
$\mu \to e\gamma$
is predicted to occur in virtually all extensions that have
been proposed.  For
example, in grand unified supersymmetric theories
\cite{barbi95}, this
upper limit increases the lower limit on the masses of the
mediating particles by 40\%, relative to the lower limit
that existed before the results of this experiment were
first announced.  Similarly, if the recently reported,
apparent deviation of
the muon anomalous magnetic moment
from the standard model prediction
\cite{gminus2}
is due to supersymmetry, this upper limit on the branching
ratio for $\mu \to e\gamma$ sets stringent limits on the
flavor-violating masses that occur in the minimal
supersymmetric standard model \cite{grae01}.

\acknowledgments

We are grateful for the support received from many LAMPF staff members and, in particular,
P.\@ Barnes, G.\@ Garvey, L.\@ Rosen and D.H.\@ White.  We wish gratefully to acknowledge
the contributions to the construction and operation of the experiment from the engineering and
technical staffs and undergraduate students at the participating institutions.  The experiment was
supported in part by the U.S.\@ Department of Energy and the National Science Foundation.

\appendix

\section{Photon Pattern Recognition}
\label{sect:photon-anal-algorithm}

\begin{figure}[tbp]
 \begin{center}
     \epsfig{file=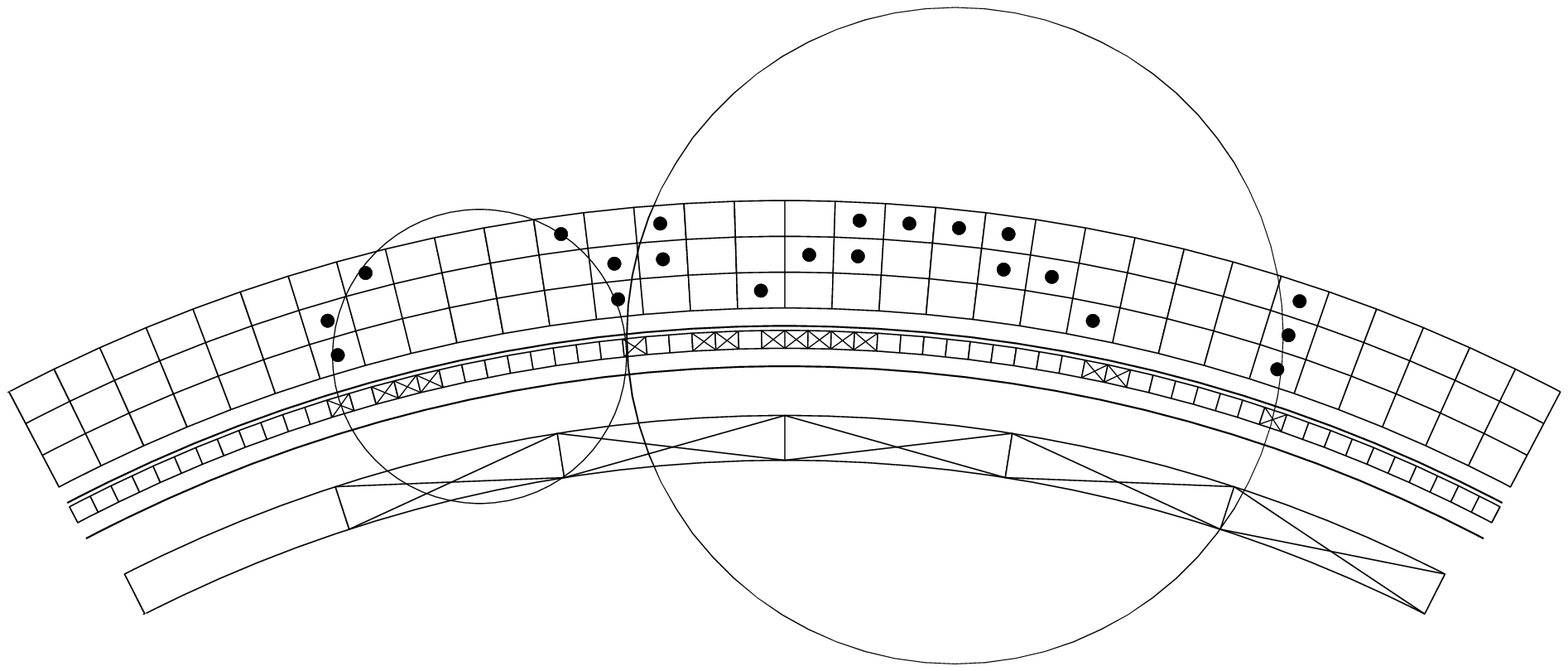,width=3.3in}
     \epsfig{file=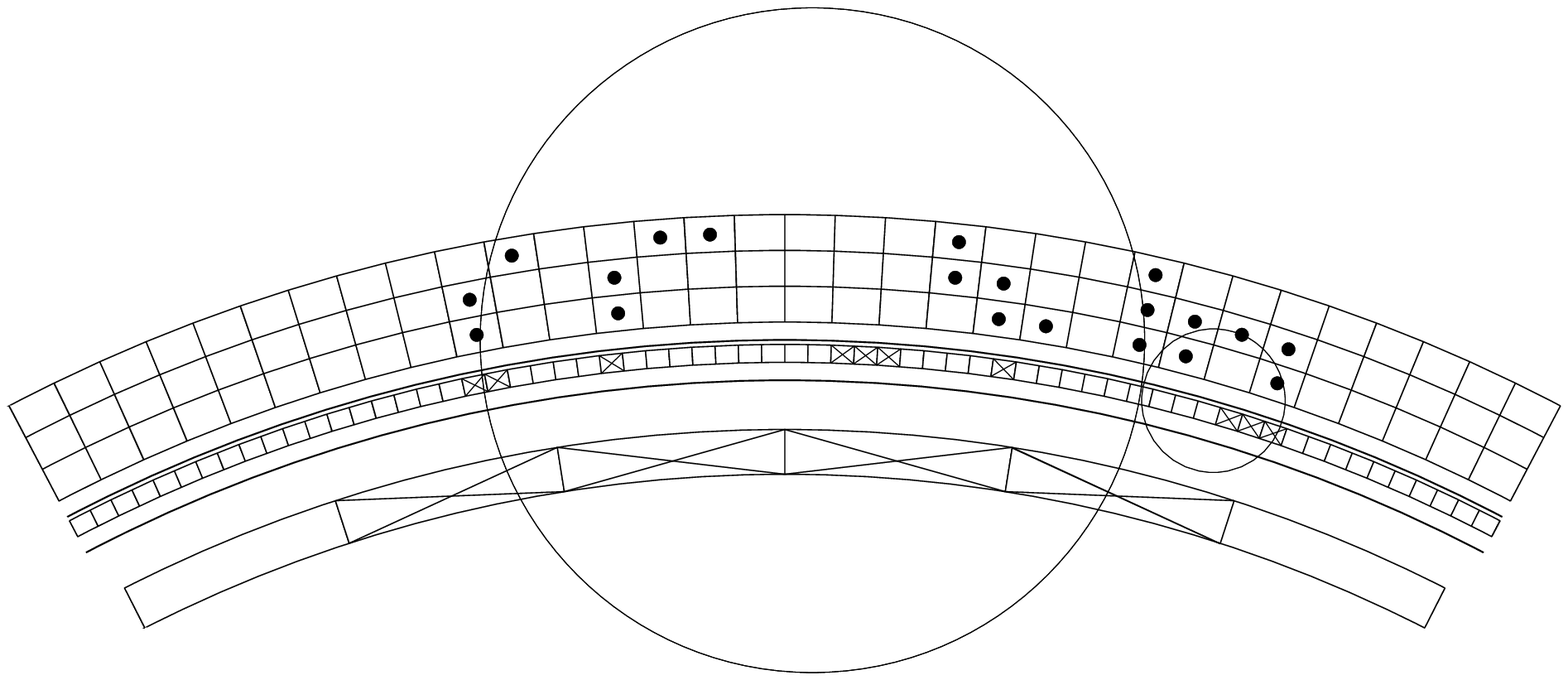,width=3.3in}
 \end{center}
  \caption{End view of two typical photon spectrometer events.  The detector elements, from
inner
  to outer radius, are plastic scintillators, lead, MWPC, lead and three layers of 
  drift chamber.  Hits in plastic scintillator and MWPC cells are shown with an 
  ``x" and hits in drift chamber cells are shown as solid circles.  In the top  
 event, both edges are classified as 3-side.  The bottom 
 event left edge corresponds to a 3-side and the right edge
 is a 2-side.  The circles shown on the figures were fits from the photon spectrometer analysis
code.  
 \label{fig:ph_pattern}} 
\end{figure}

The photon pattern recognition algorithms were optimized to find pair conversions 
due to high energy photons. Initially, active adjacent cells
within the same drift chamber cylinder were combined
into clusters of hits. If delay line information was
available, contiguous drift
chamber hits that had $z$ values differing by more than 4 cm
were assigned to separate clusters.  Pattern recognition
then began by classifying the edges of the events. To be considered as a
candidate event, at least one member of the conversion pair had to pass through all
three photon drift chambers
in a pair spectrometer during its initial arc.
Shower edges that appeared consistent with this hypothesis were
identified as ``3-side" edges. The
other member of the pair was required to pass through at least the innermost
and middle drift chambers,
but not necessarily the outermost.  Shower edges that appeared to
arise from tracks that passed through only the inner two drift chambers of a pair
spectrometer were identified as ``2-side" edges.  Typical events with 3-side and 2-side
edges are shown in Fig.\@ \ref{fig:ph_pattern}.
The 3-side edges were allowed to have a
maximum cluster width of two drift chamber cells in the first and second
drift chambers
to minimize ambiguities due to multiple passes. These constraints eliminated
some high energy photon conversions, but they reduced background and improved
resolution without significantly reducing the reconstruction efficiency. Events with
satisfactory edges were then checked for possible vertices.
For events with a 2-side edge, the vertex location was found by starting from the 2-side
edge and locating the first set of hits in the outermost drift chamber. 
Connecting this with hits in the inner two drift chambers defined the
candidate vertex location. 
Both minimum and maximum width constraints were imposed on the distance
between the 2-side edge and the vertex.  If the width constraints were not satisfied,
the cells on the 2-side were considered to be noise hits and an attempt was made to
reclassify the edge of the event as a 3-side edge.  

The on-line pattern recognition
algorithm allowed for multiple vertex candidates for events that were tagged as
having 3-sides on both edges.  Candidate vertex locations required more than
one hit in the outermost drift chamber.
This constraint eliminated most high energy Compton
scattered photons with very little loss of efficiency for 52.8 MeV pairs.
(This constraint was relaxed during the $\pi^0 \to \gamma\gamma$ studies
described in Sect.\@ \ref{sect:pi0}
for certain classes of electromagnetic
showers induced by the higher energy photon.)  A
maximum of  two contiguous cells was allowed
in the innermost drift chamber for a vertex candidate
associated with two 3-side edges.  MC simulation was used to set
additional restrictions for vertex candidates that involved non-contiguous cells in the
middle and/or the outermost drift chambers.  

The pair recognition algorithm was performed in the on-line analysis and then 
repeated during off-line processing. In the on-line code, a crude determination 
of the energy was made by estimating the transverse momentum from the width of the
event and the longitudinal momentum using the difference in the vertex and edge $z$
locations, coupled with the vertex to edge transverse distances.
The photon conversion location, ($R_{\gamma}$,$\phi_{\gamma}$,$z_{\gamma}$), was
obtained from the vertex location 
and the time of conversion, $T_{\gamma}$, was obtained from scintillators
that were hit below the edge cells.
  
In the off-line code, events were further processed by algorithms that determined those
cells that were most likely associated with the initial arc of the conversion pair. The
tagging process was repeated for each vertex candidate in an event. Tagged cells
formed the set of drift chamber hits used in the circle fits for the events. 
A non-linear least-squares fitting routine was used to obtain the best circle fits 
for the conversion pair. The $\chi^2$ that was minimized summed 
$(d_{meas}-d_{calc})^2/\sigma_d^2$ over the tagged cells
as well as $(v_{e^+}-v_{e^-})^2/\sigma_v^2$, where $d$ was the 
drift distance of each hit cell and $v_{e^+}$ ($v_{e^-}$) was the intersection point
of the positron (electron) circle and a lead conversion layer.
The choice to fit drift distances, rather than space points, eliminated the need to
resolve left-right ambiguities in the drift chambers explicitly, at the expense of
introducing additional local minima into the $\chi^2$ space.
These local minima were addressed in part by using the results of circle
fits to the $e^+$ and $e^-$ trajectories, which utilized the
left-right information, as starting points for the final non-linear fit. The vertex 
constraint in the non-linear fit
helped improve the momentum resolution of the particles
and rejected 
solutions that were not consistent with a pair 
originating from a common origin in a lead foil.
When the pattern recognition indicated that both the
$e^-$ and the $e^+$ passed through the same drift chamber cell near the event
vertex, the $\chi^2$ fit was modified to require one member of the pair to have
the observed drift distance, while the other member of the pair was required to
pass through the cell at some larger drift distance.
The code 
determined if the conversion occurred in the inner or outer lead layer by 
examining the occupancy of MWPC cells below the vertex region.  For example, 
in Fig. \@\ref{fig:ph_pattern} the conversion in the top event occurred in the inner lead 
layer, whereas the conversion in the bottom event occurred in the outer lead layer.
Additional constraints were 
placed on the circles to ensure that the fits were physically acceptable 
solutions.  Notably, circle centers were allowed to occupy only a limited region of the 
detector. 
When several vertex candidates were available, the best 
choice was assumed to be the 
one with the minimum $\chi^2$. 
%The plastic scintillators associated with the first pass were found from the 
%intersection of the circles with the scintillator barrel, and these were used 
%to determine the conversion time of the pair. 

Photon kinematic parameters were extracted from the  
best fit solution. Initial values for $p_T$ and 
the longitudinal momentum, $p_z$, were obtained 
separately for the $e^+$ and $e^-$ of the conversion pair. An average 
${dE}/{dx}$ correction for the pair was determined, based on the conversion lead 
layer.  The uncertainty in ${dE}/{dx}$ was the biggest contributor to the photon
energy resolution for conversions with dip angles less than 20$^{\circ}$. The reconstructed
momenta and the energy loss were then combined to determine a value 
for the initial photon energy. The photon angle was obtained with respect to 
the beam direction from an average of the pair angles.
Initial timing information was obtained from the scintillators 
associated with the first arcs of the conversion pair.  Corrections were made to 
account for the flight time of the positron and electron between the vertex location 
and the scintillator. The event was then checked for scintillator hits associated 
with a second pass of either member of the conversion pair. The time of the photon
decay at the target was obtained by combining scintillator times, 
corrected for the pair time-of-flight, and then correcting for the photon 
time-of-flight between the target and conversion point.

\section{ARC Orbit Calculation}
\label{app:arc}

This appendix describes the calculation of
the likely helical orbit of a positron from a $\mu \to e\gamma$ decay given the parameters
of the photon that triggered the event.  The photon reconstruction
yielded the photon ($\gamma \rightarrow
e^{+}e^{-}$) conversion location ($R_{\gamma}$,$\phi_{\gamma}$,$z_{\gamma}$) and the
calibrated TDC time ($T_{\gamma}$) for the conversion.  The photon and the positron were
assumed
to originate simultaneously from a common point in the muon stopping target and the
positron was assumed to have a momentum, {\bf p}$_e$, of 52.8 MeV/c directed opposite to that
of the photon.  From a given hit positron
scintillator and its associated Snow White triple, the positron time ($T_e$) and the spatial
location 
($x_{sw}$,$y_{sw}$,$z_{sw}$) at the termination of the orbit were obtained.  Based on this
information, ARC determined a
helical orbit along which the positron spectrometer data could be searched for hits. 

\begin{figure}[tbp]
 \begin{center}
%     \mbox{\epsfxsize=3.3in \epsfbox{fig_app1.eps}}
     \epsfig{file=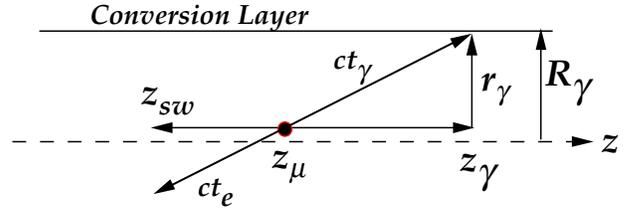,width=3.3in}
 \end{center}
\caption{Using the muon decay point in the target, Eqs.\@ \ref{eqn:app1},
   \ref{eqn:app2} and \ref{eqn:app3} follow from the figure.}
\label{fig:app1}
\end{figure}

The absolute transit times of the photon ($t_{\gamma}$) and the positron ($t_{e}$) from a
candidate $\mu \to e\gamma$ decay were not known.  However, the time difference ($T_{e} -
T_{\gamma}$) was calibrated to be equal to the difference in the transit times 
($t_{e} - t_{\gamma}$).  This identity was carefully checked on line and then validated with the 
IB data (Sect. \@\ref{sect:ib}).  The ARC algorithm calculated both $t_{e}$ 
and $t_{\gamma}$ as well as the
estimated muon decay point in the target, $x_{\mu}$, $y_{\mu}$, $z_{\mu}$, the helical orbit
radius, $R_{e}$, the helix center, $x_{c}$, $y_{c}$, and the total angular path of the helix. 
From Fig.\@ \ref{fig:app1} the following three equations can be written
for a $\mu \to e\gamma$ event:
        %c*tg/c*te = (zmu-zg)/(zsw-zmu)(1)
\begin{equation}
\frac{c  t_{\gamma}}{c t_{e}}=
\frac{ z_{\mu}-z_{\gamma} } { z_{sw}-z_{\mu} };
\label{eqn:app1}
\end{equation}
         %c*teg = c*te - c*tg(2)
\begin{equation}
c\, (T_{e} -T_{\gamma}) = c   t_{e} - c   t_{\gamma};
\label{eqn:app2}                             
\end{equation}
        %(c*tg)^2 = Rg^2 + (zmu-zg)^2(3)
\begin{equation}
(c  t_{\gamma})^{2}=r_{\gamma}^{2} + \left( z_{\mu}-z_{\gamma} \right) ^{2}.
\label{eqn:app3}
\end{equation}
The initial approximation for $r_{\gamma}$ was $R_{\gamma}$, the 
radius of the photon conversion layer. 
Equations \ref{eqn:app1} - \ref{eqn:app3} 
were solved by iteration for  
$t_{\gamma}$, $t_{e}$, and $z_{\mu}$. Only two
iterations were required for sufficient accuracy.

\begin{figure}[tbp]
 \begin{center}
     \epsfig{file=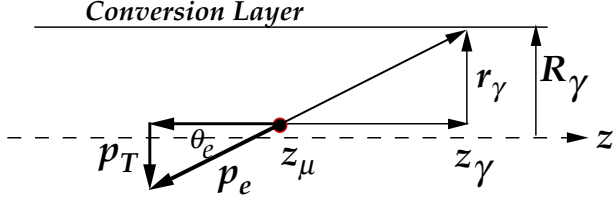,width=3.3in}
 \end{center}
\caption{The momentum of the positron is related to the orbit radius for the positron
     orbit, leading to Eq.\@ \ref{eqn:app5}}
\label{fig:app2}
\end{figure}

From Fig.\@ \ref{fig:app2}, the orbit radius, $R_{e}$ was calculated from  
        %p_perp = pe * sin(theta_e)(4)
%\begin{equation}
%p_{T}=p_{e} \, \sin \left( \theta_{e} \right)
%=p_{e}\,  \frac{r_{\gamma}}{ \sqrt{ r_{\gamma}^{2}+ 
%\left( z_{\gamma}-z_{\mu} \right) ^{2} }},
%\label{eqn:app4}
%\end{equation}
%   %               = pe * Rg/sqrt(Rg^2+(zg-zmu)^2)(5)
%\begin{equation}
%p_{T};
%\label{eqn:app5}
%\end{equation}
          %Re = p_perp/qB(6)
\begin{equation}
R_{e}= \frac{p_{T}}{q \space  B}
= \frac{p_{e}}{qB} \, 
\frac{r_{\gamma}}{ \sqrt{ r_{\gamma}^{2}+
\left( z_{\gamma}-z_{\mu} \right) ^{2} }}.
\label{eqn:app5}
\end{equation}
Here $q$ is the positron charge and $B$ is the magnetic field.
Given $z_{\gamma}$, $z_{\mu}$, and
$r_{\gamma}$ $\approx$ $R_{\gamma}$, an estimate of the orbit
radius, $R_{e}$, was obtained.  The quotient $p_{e}/qB$ = 11.73 cm represents the
largest possible orbit radius for the positron in the 1.5 T magnetic field.  

\begin{figure}[tbp]
 \begin{center}
     \epsfig{file=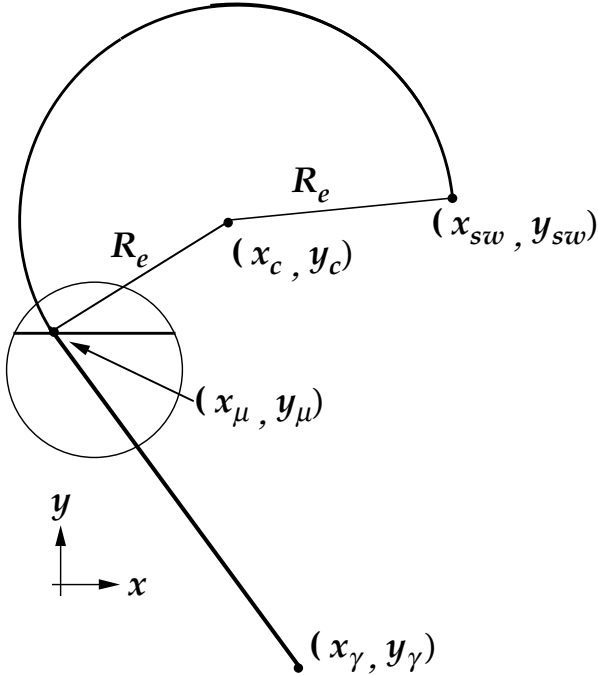,width=3.3in}
 \end{center}
\caption{This view along the axis of the positron helix leads to a determination of the 
     center of the helical orbit and the $x$-coordinate of the muon decay obtained from
     Eqs.\@ \ref{eqn:app6},  \ref{eqn:app7} and \ref{eqn:app8}}
\label{fig:app3}
\end{figure}

The planar target sloped in the $y$-$z$ plane, allowing 
$y_{\mu}$ to be determined from $z_{\mu}$.
Using Fig.\@ \ref{fig:app3}, the coordinates of the axis of the helical orbit
($x_{c}$,$y_{c}$) and the $x$-coordinate of the muon decay in the muon stopping
target ($x_{\mu}$) were calculated from the following three equations:
      %(yg - ymu)/(xg - xmu) = (xmu - xc)/(ymu - mc)(9)
\begin{equation}
- \frac{ y_{\gamma}-y_{\mu}}
{ x_{\gamma}-x_{\mu}}=
\frac{ x_{\mu}-x_{c}}
{ y_{\mu}-y_{c} };
\label{eqn:app6}
\end{equation}
       %(xsc - xc)^2 + (ysc - yc)^2 = Re^2(10)
\begin{equation}
 \left( x_{sw}-x_{c} \right) ^{2} +  
\left( y_{sw}-y_{c} \right) ^{2}=R_{e}^{2};
\label{eqn:app7}
\end{equation}
       %(xmu - xc)^2 + (ymu - yc)^2 = Re^2 .(11)
\begin{equation}
 \left( x_{\mu}-x_{c} \right) ^{2} +  
\left( y_{\mu}-y_{c} \right) ^{2}=R_{e}^{2}.
\label{eqn:app8}
\end{equation}
With ($x_{\mu}$,$y_{\mu}$,$z_{\mu}$)
determined, the value for
$r_{\gamma}$ was calculated.
The entire process was repeated using these values to obtain better estimates of $t_{\gamma}$,
$t_{e}$, $x_{\mu}$, $y_{\mu}$, $z_{\mu}$, $R_{e}$, $x_{c}$, and $y_{c}$,

Employing the coordinates of the beginning ($x_{\mu}$,$y_{\mu}$,$z_{\mu}$), 
ending ($x_{sw}$,$y_{sw}$,$z_{sw}$), and axis ($x_{c}$,$y_{c}$) of 
the helical positron orbit, the helix rotation angle
($\theta_{e \mu}$) was calculated from ($x_{\mu}$,$y_{\mu}$) to ($x_{sw}$,$y_{sw}$).
The total angle traversed by the positron can be written as 
      %n_loop = (1/2pi) * [(qB*te/m) - th_emu].(13)
\begin{equation}
\theta_{e\mu} +2 \space  \pi \space  n_{loop} = 
\left( \frac{q B t_{e}}{m} \right), 
\label{eqn:app13}
\end{equation}
where $m$ is the mass of the positron and $n_{loop}$ is the number of loops.  
The nearest integer value for $n_{loop}$ was determined from  Eq.\@ \ref{eqn:app13} and 
the integer result was substituted back in  Eq.\@ \ref{eqn:app13} to obtain a better
estimate of $t_e$.
The new value of $t_{e}$ was used
in Eq.\@ \ref{eqn:app1} to improve the orbit
parameters further.
With the orbit parameters known, 
the spatial coordinates where the orbit
intersected the positron MWPCs were calculated.

\section{Positron Pattern Recognition}
\label{app:pos-pattern}

%\subsubsection{Space Points}

The principle job of the positron pattern recognition 
code was to find the space points, called clusters,
where positrons crossed the MWPCs and then map the clusters into helical tracks.  
Clusters were defined as nearly contiguous groups of hits in either the anodes or
cathodes that were normally associated with the response of a chamber to a track
crossing.  
%Wire or cathode stripe hits came in three types:  a hit wire, an unhit wire,
%or a dead wire.  
The clustering algorithm took
into consideration the imperfections of the detector (about 7\% dead wires and cathode stripes
and inefficiencies in the chambers), as well as the
dependence of
cluster widths on the entry angle of the track with the chamber.
The anodes and cathodes were treated slightly differently because of the 
differences in their modes of signal generation.  Track reconstruction efficiency
was mostly insensitive to the choice of clustering algorithm, but the method that was selected
optimized the resolutions; the projection to the decay point in the target was more sensitive than
the energy.  For high rate data, clustering was only applied to chambers 
that were potentially
involved in a signal event.

\begin{figure}[tbp]
   \begin{center}
     \mbox{\epsfxsize=3.3in \epsfbox{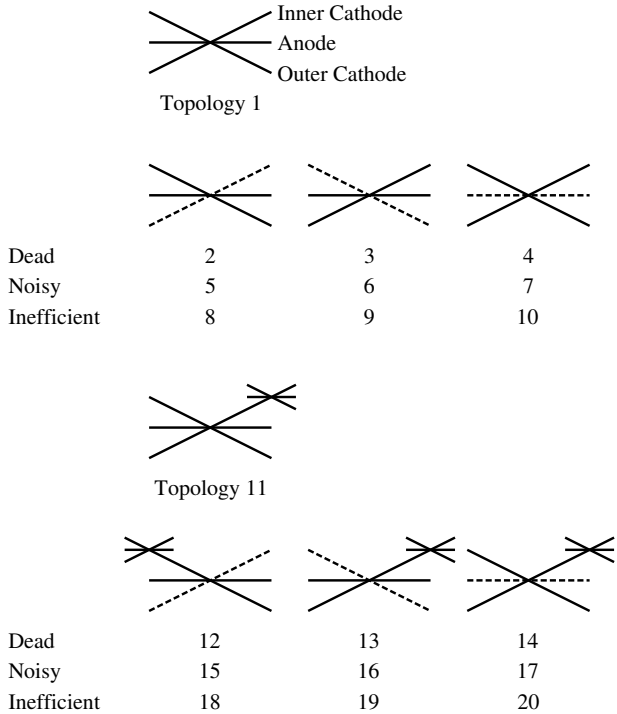}}
   \end{center}                                                  
   \caption{The different types of space point topologies.
The solid lines represent clean hits.  Illustrations with
dashed lines represent three distinct topologies, where
the dashed line may be replaced by a dead channel, a
noisy cluster, or a known inefficiency.  They were
assigned distinct numbers and were studied separately
during the development of the pattern recognition
program.  Additional topologies were also defined and
studied, such as triples with two noisy clusters, but were
not used in the final analysis.}
   \label{fig:5b1}
\end{figure}

The term ``double'' describes the 
spatial overlap of an anode cluster with a cluster from one of the foils.  A triple
required the z coordinate of two doubles associated with a single anode be
the same to within 1 cm.  If a cathode cluster contributed to more than one 
double, only the two triples with the smallest differences in anode crossings were 
retained.  Anode-only clusters were allowed in the Snow White chamber when
positron tracks were leaving the chamber toward larger radial distance.  
Triples and doubles were of
varying quality and were denoted according to their ``crossing topologies''.  The types
of crossing topologies are displayed in Fig.\@ \ref{fig:5b1}.
Noisy clusters had an excessive number of wires;
e.g., for the anodes, a noisy cluster in a dwarf contained more than 8 wires. 
Clusters in regions of known inefficiencies were noted.  Clusters that could have been
associated with another crossing were tagged, since it was possible for
triples and doubles to be formed from random hits at high rates.

The high rate
code only examined crossing topologies if the anode cluster was within the end-view
ARC windows.  
Only crossing topologies 1, 2-7, and 11 were used as parts
of tracks.  The coordinates of the intersections were calculated from the
centers of the cluster crossings.  In the case of differing widths for inner and outer
cathode clusters, the narrower width was used to determine the coordinate.

The intrinsic resolutions for a normal crossing were 0.065 cm rms for the anodes
and 0.29 cm rms for the cathodes.  The weights (1/$ \sigma ^2$) for fitting were assigned
to the space points to reflect
an intrinsic resolution, the entrance angle, and the differences in the cathode cluster
widths.  These weights, as a function of
entrance angle and cluster widths, were in good agreement with the residuals from
track fitting.

%\subsubsection{Tracks}

As an initial step in obtaining tracks, all combinations of three clean (e.g., $<$\,8
wires) anode-wire clusters were used to extract the parameters of circles that passed
through these anodes.  For high rate data, the combinations were restricted to groups
of three clean anodes in distinct ARC windows.  All wires (ignoring ARC windows)
in the involved chambers were examined for anodes within a cord length of 0.75 cm,
and successes were associated with this circle.  The circle
parameters were extracted by fitting these data.  After all possible combinations were fit, the
parameters were compared to eliminate duplicates.

There could be considerable ambiguity with respect to the number of loops a real track
made for a given circle due to inefficiencies and accidental
degeneracies of the hits.  At high rates, the 
large number of hits resulted in many possibilities, some of which were artificial.
Starting with circles, helical track fragments were sought by looking for 
straight lines in the unrolled system.  Combinations of track fragments were
compared to determine if they fell on the same straight line where the 2$ \pi$ ambiguity
around the circle was considered.  If enough fragments lay on the same straight
line for a particular set of slopes and intercepts, a trial track was identified.

The trial track was then extended in both directions toward the target and toward
the scintillators.  If the extensions passed sufficiently close to triples having the
correct topologies, these hits were added to the trial tracks.  On the target side, there
was an ambiguity as to the origin of the track, so a minimum criterion for the number
of triples and doubles in the first loop was applied.  On the scintillator side, a final
triple in the Snow White chamber was required for a track to be retained.  Roughly, for a
candidate track to be kept, it must retain 60-70\% of the possible hits.

Non-uniformities in the magnetic field led to a complexity
in the above procedure for events with a high number of loops.
Tracks still projected as circles but
they deviated from a straight line in the unrolled view \cite{jackson}.
In order to keep these
tracks, the extrapolation of the fragment slopes had to account for the deviations that
increased the longitudinal distance traversed per loop as the field
weakened.  The field non-uniformities were small and well described by a
polynomial; the ratio of the first and second terms to the 0$^{th}$ term were
$-1.8 \times 10^{-4}$ cm$^{-1}$ and $-7.1 \times 10^{-6}$ cm$^{-2}$, 
respectively, over the range of
$\pm$70 cm.  Though the terms were small, the corrections were many centimeters for
an event making many loops.

Once the hits were identified, the parameters characterizing the track, the
circumference around the chamber and the longitudinal position including the
$z$-correction term, were extracted via a non-linear least squares fit.  The
circumference was chosen because the hits were confined to lie on the circles that
were defined by the chamber wires.  The parameters were defined by the helix
equations
\begin{eqnarray}
x(\phi) & = & x_c + \rho \cos (\phi + \phi_0), \nonumber \\
y(\phi) & = & y_c + \rho \sin (\phi + \phi_0), \\
z(\phi) & = & z_0 + (\rho \tan \lambda) \phi \nonumber .
\end{eqnarray}
The derivatives of the predicted measurements with respect to the parameters
that were needed to make up the curvature matrix of the least squares fit 
are analytic,
making the fitting process reasonably fast \cite{stantz97}.

However, using this procedure it was possible to find the same track many
times and, at high rates, to find false tracks.  
A number of criteria were used to sort the best 
candidates from the list to maximize the probability of finding only real tracks once.  For
multiple
tracks associated with the same circle, the track 
with the greatest fractional occupancy was retained, and tracks with the same slope and 
center were concatenated to a single track.  Tracks were required to have a certain 
minimum number of triples and doubles and the cutoffs depended on the number of loops
and the number of dwarfs involved.  
%Finally, each track found was required to
%have a $ \chi^2$ per degree of freedom below 3.0 from the final fit, which is described
%below.

\end{document}